\documentclass[twocolumn]{aastex631}

\shortauthors{Lewin et al.}
\graphicspath{{./}{/}}

\begin{document}

\title{X-ray Reverberation Mapping of Ark~564 using Gaussian Process Regression}
\correspondingauthor{Collin Lewin}
\email{clewin@mit.edu}

\author[0000-0002-8671-1190]{Collin Lewin}
\affiliation{MIT Kavli Institute for Astrophysics and Space Research, MIT, 77 Massachusetts Avenue, Cambridge, MA 02139, USA}

\author[0000-0003-0172-0854]{Erin Kara}
\affiliation{MIT Kavli Institute for Astrophysics and Space Research, MIT, 77 Massachusetts Avenue, Cambridge, MA 02139, USA}

\author[0000-0002-4794-5998]{Dan Wilkins}
\affiliation{Kavli Institute for Particle Astrophysics and Cosmology, Stanford University, 452 Lomita Mall, Stanford, CA 94305, USA}

\author[0000-0003-4216-7936]{Guglielmo Mastroserio}
\affiliation{Cahill Center for Astronomy and Astrophysics, California Institute of Technology, 1200 California Boulevard, Pasadena, CA 91125, USA}

\author[0000-0003-3828-2448]{Javier A. Garc\'ia}
\affiliation{Cahill Center for Astronomy and Astrophysics, California Institute of Technology, 1200 California Boulevard, Pasadena, CA 91125, USA}
\affiliation{Dr. Karl Remeis-Observatory and Erlangen Centre for Astroparticle Physics
Friedrich-Alexander-Universit\"at Erlangen-N\"urnberg
Sternwartstr. 7
96049 Bamberg
Germany}

\author[0000-0002-2905-9239]{Rachel C. Zhang}
\affiliation{MIT Kavli Institute for Astrophysics and Space Research, MIT, 77 Massachusetts Avenue, Cambridge, MA 02139, USA}
\affiliation{Department of Physics and Astronomy and Center for Interdisciplinary Exploration and Research in Astrophysics (CIERA),
Northwestern University, 2145 Sheridan Road, Evanston, IL 60208, USA}

\author[0000-0003-2658-6559]{William N. Alston}
\affiliation{European Space Agency (ESA), European Space Astronomy Centre (ESAC), Villanueva de la Canada, Madrid, E-28691, Spain}

\author[0000-0002-8908-759X]{Riley Connors}
\affiliation{Cahill Center for Astronomy and Astrophysics, California Institute of Technology, 1200 California Boulevard, Pasadena, CA 91125, USA}

\author[0000-0003-4583-9048]{Thomas Dauser}
\affiliation{Dr. Karl Remeis-Observatory and Erlangen Centre for Astroparticle Physics
Friedrich-Alexander-Universit\"at Erlangen-N\"urnberg
Sternwartstr. 7
96049 Bamberg
Germany}

\author[0000-0002-9378-4072]{Andrew Fabian}
\affiliation{Institute of Astronomy, Madingley Road, Cambridge, CB3 0HA, UK}

\author[0000-0002-5311-9078]{Adam Ingram}
\affiliation{Department of Physics, Astrophysics, University of Oxford, Denys Wilkinson Building, Keble Road, Oxford OX1 3RH, UK
}
\affiliation{School of Mathematics, Statistics and Physics, Newcastle University, Herschel Building, Newcastle upon Tyne, NE1 7RU, UK}

\author[0000-0002-9639-4352]{Jiachen Jiang}
\affiliation{Institute of Astronomy, Madingley Road, Cambridge, CB3 0HA, UK}

\author{Anne Lohfink}
\affiliation{Department of Physics, Montana State University, P.O. Box 173840, Bozeman, MT 59717-3840, USA}

\author[0000-0002-2235-3347]{Matteo Lucchini}
\affiliation{MIT Kavli Institute for Astrophysics and Space Research, MIT, 77 Massachusetts Avenue, Cambridge, MA 02139, USA}

\author[0000-0002-1510-4860]{Christopher S. Reynolds}
\affiliation{Institute of Astronomy, Madingley Road, Cambridge, CB3 0HA, UK}

\author[0000-0002-6562-8654]{Francesco Tombesi}
\affiliation{Astrophysics Science Division, NASA/Goddard Space Flight Center, Greenbelt, MD 20771}
\affiliation{Department of Astronomy, University of Maryland, College Park, MD 20742}
\affiliation{Department of Physics, University of Rome “Tor Vergata”, Via della Ricerca Scientifica 1, I-00133 Rome, Italy}
\affiliation{INAF Astronomical Observatory of Rome, Via Frascati 33, 00078 Monteporzio Catone (Rome), Italy}

\author[0000-0003-0070-9872]{Michiel van der Klis}
\affiliation{Astronomical Institute, Anton Pannekoek, University of Amsterdam, Science Park 904, NL-1098 XH Amsterdam, Netherlands}

\author[0000-0002-1742-2125]{Jingyi Wang}
\affiliation{MIT Kavli Institute for Astrophysics and Space Research, MIT, 77 Massachusetts Avenue, Cambridge, MA 02139, USA}

\begin{abstract}
  Ark~564 is an extreme high-Eddington Narrow-line Seyfert~1 galaxy, known for being one of the brightest, most rapidly variable soft X-ray AGN, and for having one of the lowest temperature coronae. Here we present a 410~ks NuSTAR observation and two 115~ks XMM-Newton observations of this unique source, which reveal a very strong, relativistically broadened iron line. We compute the Fourier-resolved time lags by first using Gaussian processes to interpolate the NuSTAR gaps, implementing the first employment of multi-task learning for application in AGN timing. By fitting simultaneously the time lags and the flux spectra with the relativistic reverberation model {\sc reltrans}, we constrain the mass at $2.3^{+2.6}_{-1.3} \times 10^6M_\odot$, although additional components are required to describe the prominent soft excess in this source. These results motivate future combinations of machine learning, Fourier-resolved timing, and the development of reverberation models.  

\end{abstract}

\section{Introduction} \label{sec:intro}
X-ray emission from Active Galactic Nuclei (AGN) is powered by accretion onto a central supermassive black hole. The accretion disk emits thermally, producing UV seed photons that are Compton up-scattered to X-ray energies by a region of high-energy particles located close to the black hole known as the corona \citep{1979Natur.279..506S}. The Comptonized emission from the corona creates a direct coronal continuum in the X-ray spectrum. This emission then irradiates and is reprocessed by the inner accretion disk, producing correlated variability that is delayed on the order of the corona-to-disk light travel time (see \citealt{2014A&ARv..22...72U, 2021iSci...24j2557C} for a recent review). This reprocessing produces the reflection component of the spectrum, including the iron K emission lines around 6.4--6.9~keV, the iron K-edge around 7--8~keV, the broad Compton hump peaking near 20~keV, and a collection of fluorescent lines from iron and lower Z elements at softer energies \citep{1991MNRAS.249..352G, 2005MNRAS.358..211R}. 

In addition to the flux spectrum, the reflection features arise when plotting the time delays between correlated variability in each energy band to a common reference band (the lag-energy spectrum) \citep[e.g.][]{2009Natur.459..540F, 2011MNRAS.412...59Z, 2016MNRAS.462..511K}. Features in the lag-energy spectrum associated with reverberation, namely from the soft excess \citep{2013MNRAS.431.2441D} and the iron K$\alpha$ line \citep{2016MNRAS.462..511K}, have been detected in over 20 sources and have unlocked a new approach for probing AGN accretion flows. 

In both the flux spectrum and the lag-energy spectrum, the shape and prominence of the relativistic reflection features vary significantly with the system's geometry and physical properties \citep{2000PASP..112.1145F, 2014MNRAS.438.2980C}. This has motivated increasingly sophisticated models to describe the reflection features observed in both the types of spectra.  \cite{2014MNRAS.438.2980C} first fit the Fe K lags observed in NGC 4151 by calculating general relativistic transfer functions for reverberation for 28 combinations of height, spin, inclination, and reflection fraction. Since then, models have developed to placing constraints on 7 physical parameters of Mrk 335 by simultaneously modeling both the flux spectrum and lag-energy spectra in two frequency ranges \citep{2020MNRAS.498.4971M}. These data were fit using the same model used in this paper, the relativistic reverberation mapping model \textsc{reltrans}, which models the direct coronal continuum as well as the reflection spectrum and time lags \citep{reltrans, 2021MNRAS.507...55M}. The model also showed success in describing the time lags observed in 5 epochs during the hard-to-soft state transition of black hole X-ray binary MAXI J1820+070 by \cite{2021ApJ...910L...3W} who, from modeling the reverberation, revealed that the corona extends vertically during the transition. Several previous studies \citep[e.g.][]{2020MNRAS.498.4971M, 2020ApJ...893...97Z, 2021ApJ...910L...3W}, however, find discrepancies in the physics, namely the height of the corona, inferred from modeling the time lags versus the flux spectra. 

In this paper, we model the X-ray spectrum and time lags of well-known Narrow-line Seyfert I (NLS1) Ark~564 by simultaneously fitting both the NuSTAR and XMM-Newton flux spectra and lag-energy spectra in three frequency ranges. Doing so with the same reverberation model allows us to probe model consistency and properties of the system by comparing how a single description for the accretion flow and central engine describes the flux spectra versus the time lags.

Most previous work on Ark~564 has focused on modeling the source's flux spectrum \citep[e.g.][]{2015A&A...577A...8G,  Kara_2017, density}, which exhibits features typical for a NLS1, including a strong soft excess and a steep spectral slope \citep{10.1093/mnras/285.3.L25}. The first deep observation of the source with NuSTAR observatory ($\sim$200~ks) revealed that Ark~564 also has one of the coolest coronas discovered to-date, with an electron temperature of roughly 15~keV \citep{Kara_2017}. 

The Fourier frequency-resolved timing approach has been fruitful in understanding Ark~564 and other sources by isolating the time lags from variability occurring on different timescales (i.e. frequencies) due to distinct physical processes. Ark~564's high-frequency soft lags were first discovered tentatively by \cite{2007MNRAS.382..985M}. The soft lags were then confirmed by \cite{2012ApJ...760...73L} and then \cite{2013MNRAS.434.1129K}, who also found evidence for the Fe K lag. The lowest frequency one can probe with this approach, however, is limited by the length of one’s longest continuous observation--a significant limit for data collected with instruments in low-Earth orbit. Due to NuSTAR's data gaps from Earth occultations, however, we cannot probe long enough timescales to access the low-frequency hard lags, thought to be associated with the inward propagation of accretion rate fluctuations \citep{2001MNRAS.327..799K, 2006MNRAS.367..801A}. 

\begin{figure*}[ht!]
    \centering
    \includegraphics[width=0.85\textwidth]{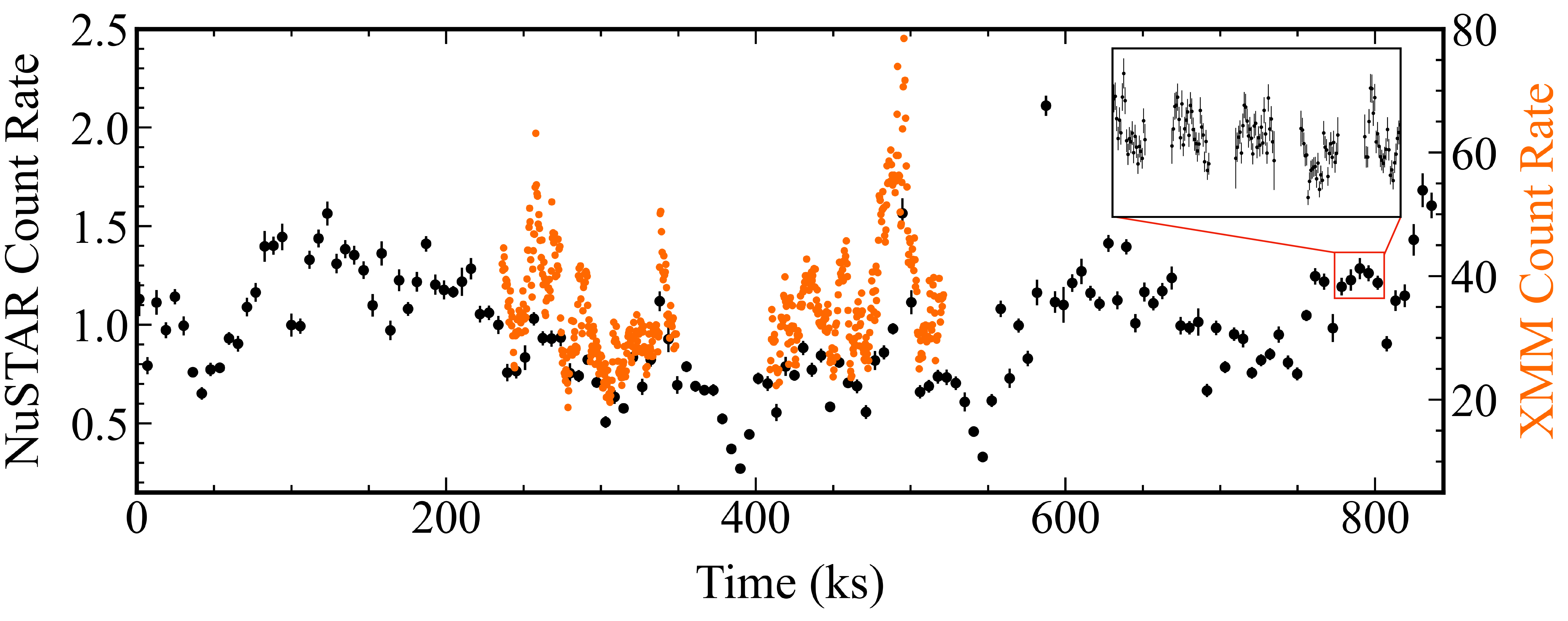}
    \caption{Background-subtracted broadband NuSTAR FPMA+FPMB lightcurves (black; left axis) for the new 410~ks observation---with each point representing data averaged over single orbit---and two partially-simultaneous XMM-Newton observations (orange; right axis) with an exposure totalling 230~ks in 400s bins. The inlay shows the data for five NuSTAR orbits with less coarse binning, including the low-Earth orbit data gaps motivating use of Gaussian process regression. The source is highly variable, with multiple flares occurring throughout the observations.}
    \label{fig:lcs}
\end{figure*}

In order to probe Ark~564's low-frequency hard lags, \cite{2012ApJ...760...73L} corrected for the breaks between XMM-Newton observations by applying the maximum-likelihood method of \cite{2010MNRAS.403..196M}, which models the autocorrelations and cross-correlation (and, thus, the frequency-dependent time lags) by fitting the lightcurves. This approach was implemented by \cite{2013ApJ...777...24Z}, who aimed to extend reverberation studies to the non-continuously sampled data (beyond XMM-Newton observations) due to Earth occultations. \cite{Kara_2017} applied the maximum-likelihood method of \cite{2013ApJ...777...24Z} in order to access the NuSTAR low-frequency lags, but were unable to probe the reverberation lags at high-frequencies due to limited statistics. 

More recently, Gaussian process regression (GPR) has grown in popularity and has shown success in modeling the underlying probability distribution of sparse lightcurves of asteroids \citep{2021arXiv211112596W}, AGN \citep{2014ApJ...788...33K, Wilkins_2019, Griffiths_2021}, and stars \citep{2009MNRAS.395.2226B, 2017ApJ...840...49C,  2017MNRAS.464.1353M}. Continuous realizations of the lightcurve can then be generated after modeling the observed variability, allowing for a probabilistic treatment of data generated in the gaps. Using this method, \citet{Wilkins_2019} recovers a simulated lag of 200s from $5 \times 10^{-4}$~Hz and $ 2 \times 10^{-3}$~Hz within a fractional error of 2\% from simulated \textit{NICER} observations of Ark~564 with low-Earth orbit data gaps. Unlike the approaches of \cite{2010MNRAS.403..196M, 2013ApJ...777...24Z}, Gaussian processes do not require any model assumptions for the cross-correlation between energy bands---the models describing the variability in each lightcurve produce this on their own. The lightcurve in each energy band is treated as an independent entity, in the same way as when applying traditional Fourier techniques.

We present three new observations of Ark~564: two observations with XMM-Newton totalling 230~ks that are simultaneous with a longer exposure of 410~ks with NuSTAR, the latter being the longest single exposure of the observatory to-date. Thanks to these new observations and the implementation of GPR to maximize use of the NuSTAR data, we are able to probe the time lags in 3 distinct frequency ranges for both instruments spanning 0.01--0.9~mHz. The ability to simultaneously model six lag-energy spectra and the flux spectra is a combination of new data and the flexibility of the \textsc{reltrans} model. 

The paper is organized as follows: our observations and data reduction are presented in Section \ref{sec:reduc}. An introduction to Gaussian processes, multi-task learning, and kernel functions for regression application to NuSTAR data is provided in Section \ref{sec:gps}. The basics of performing Fourier-resolved timing analysis on our GPR results are outlined in Section \ref{sec:fourier}. We present the time lags and spectra, and physical constraints from fitting with a reverberation model in Section \ref{sec:results}, which are discussed further in Section \ref{sec:discussion}.

\begin{figure*}[ht!]
    \centering
    \includegraphics[width=0.73\textwidth]{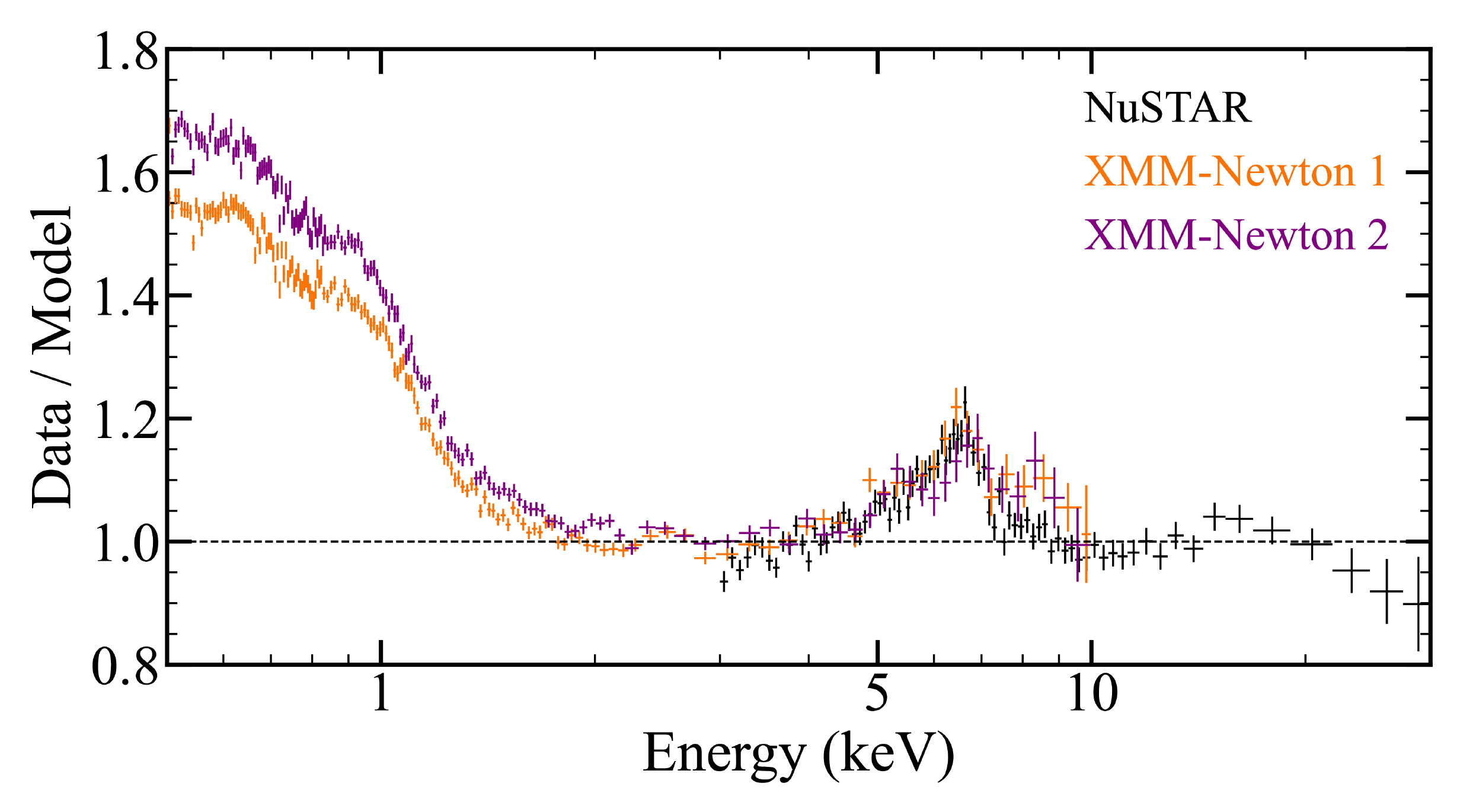}
    \caption{NuSTAR spectra (black) from the new 410~ks observation and the XMM-Newton spectra from the first and second 115~ks observations (orange and purple, respectively) fit to a simple, steep ($\Gamma = 2.57$) power-law with Galactic absorption.}
    \label{fig:obs_spec}
\end{figure*}

\section{Observations and Data Reduction} \label{sec:reduc}
\begin{deluxetable}{cccc}[t!]
\tablecaption{The observations used for this analysis, with columns indicating the observatory, observation ID, start date of the observation, and exposure time. \label{tab:obsids}}
\tablehead{\colhead{Observatory} & \colhead{Obs. ID} & \colhead{Obs. Date} & \colhead{Exposure (s)}}
\startdata
NuSTAR & 60101031002 & 2015-05-22 & 211209 \\
 & 60401031002 & 2018-06-09 & 38090 \\
 & 60401031004 & 2018-11-28 & 408958 \\
 XMM-Newton & 0830540101 & 2018-12-01 & 114900	\\
 & 0830540201 & 2018-12-03 & 114400	
\enddata
\end{deluxetable}

Our analysis uses a combination of new and archival data of Ark~564 from NuSTAR and XMM-Newton observatories, which are shown in Table \ref{tab:obsids}. We present 3 new observations of Ark~564: 410~ks by NuSTAR (PI: E. Kara)--the observatory's longest exposure yet--collected beginning 2018 November 28, as well as two 115~ks observations collected by XMM-Newton in early December 2018 (PI: E. Kara), partially simultaneous with the new NuSTAR observation. Lightcurves of these observations are shown in Figure \ref{fig:lcs}, which show the source's high variability. The 2--10~keV flux of these new XMM-Newton observations is $1.4 \times 10^{-11}$ erg/cm$^2$/s, and a 10--20~keV flux of $3.6 \times 10^{-12}$ erg/cm$^2$/s for the new NuSTAR observation. In addition to our new observations, we use all of the archival NuSTAR data for a total of 660~ks from NuSTAR: 210~ks collected on 2015 May 22, during which a $\sim$5~ks flare was observed \citep{Kara_2017}, and 38~ks of data collected on 2018 June 9.

The NuSTAR data was reduced using the NuSTAR pipeline (\textsc{nupipeline}). This consists of processing Level 1 data products with the NuSTAR Data Analysis Software (\textsc{nustardas v2.0.0}) and then creating and calibrating the Level 2 event files with CALDB version 20201130. We use a circular source region with a 50-arcsec radius and a circular background region with a 60-arcsec radius for both Focal Plane Module A and Focal Plane Module B (FPMA, FPMB) instruments. The FPMA and FPMB lightcurves are binned to 128-second bins and then added after background subtraction to improve signal-to-noise. 

We find the NuSTAR spectra to be consistent (within a maximal deviation of 10 percent) across observations and thus combine spectra, resulting in a single spectrum per FPM. The energy bins of the spectra are grouped in order to oversample the instrument's resolution by a factor of 3 and to have a signal-to-noise of at least 3$\sigma$ per bin. 

For the XMM-Newton data, we focus spectral analysis on data from the EPIC-pn instrument \citep{epicpn} due to the instrument's larger effective area. We reduced the EPIC-pn data using the XMM-Newton Science Analysis System (\textsc{sas v18.0.0}). We avoid background flares by constructing a good time interval filter using a background count rate cutoff of 0.4 counts/s and avoid spurious detections using the conditions PATTERN$\leq$4 and FLAG$==$0. We produce our spectra and lightcurves using circular source and background regions with a radius of 35-arcsec. The lightcurves were extracted and corrected using the tools {\sc evselect} and {\sc epiclccorr} with a binning to 10-second bins. The spectra were extracted using the tools {\sc evselect} and {\sc backscale}, with the redistribution matrices (RMF) and ancillary response files (ARF) being generated for each observation using the {\sc rmfgen} and {\sc arfgen} tools. We evaluated the amount of pile-up per observation using the SAS task {\sc epatplot}. Minor pile-up does appear to be present in the XMM-Newton observations, particularly below 0.5~keV. As such, we do not fit below 0.5~keV in the flux spectra and time lags throughout our analysis. As a check, we find the spectra and time lags to be very consistent (well within error) when using an annular source region in attempt to improve the pile-up versus a circular source region.

We also find an effect occurring above 7~keV, where there is an observed excess in both single and double events. The XMM-Newton spectra exceeds the NuSTAR spectra at these energies by $\sim$20\%, similar to the inconsistency between spectra reported by \cite{Kara_2017}. We find the effect does not improve when excising the bright center of the source with an annular source region. The discrepancy improves to $\sim$10\% when implementing the XMM-Newton current calibration file (CCF) released on 2022 April 7\footnote{CCF Release Note: \url{https://xmmweb.esac.esa.int/docs/documents/CAL-SRN-0388-1-4.pdf}}, which includes the \texttt{ABSCORRAREA} correction to the effective area above roughly 4~keV, but most significantly affects 7--10~keV. Figure \ref{fig:7_10_issue} shows the discrepancy before and after applying the correction to the effective area. We use the data that includes this correction for the entirety of our analysis. We note that, because of this discrepancy between XMM-Newton and NuSTAR, the equivalent width of the iron line is greater in NuSTAR than in XMM-Newton by a factor of $\sim$2 without the correction. We caution investigators modelling just the XMM-Newton spectra, as the inclusion of this 7--10~keV band will result in an artificially weak iron line.
 
Just as we did with the NuSTAR spectra, we combine the XMM-Newton spectra across observations after finding the maximum deviation of the spectra to be consistent within 15 percent. The spectra are binned in order to oversample the instrument's resolution by a factor of 3 and to have a signal-to-noise of at least 3$\sigma$ per bin. 

The ratio of the spectra to a simple power-law model is shown in Figure \ref{fig:obs_spec}. Ark~564 shows a steep spectral slope as well as strong features associated with relativistic reflection, including a strong soft excess below 1~keV. The cut-off energy appears to be low, and the prominent iron K feature is broad, both in agreement with \cite{Kara_2017}.

\section{Gaussian Processes} \label{sec:gps}
 \begin{figure}[t!]
    \centering
    \includegraphics[width=\linewidth]{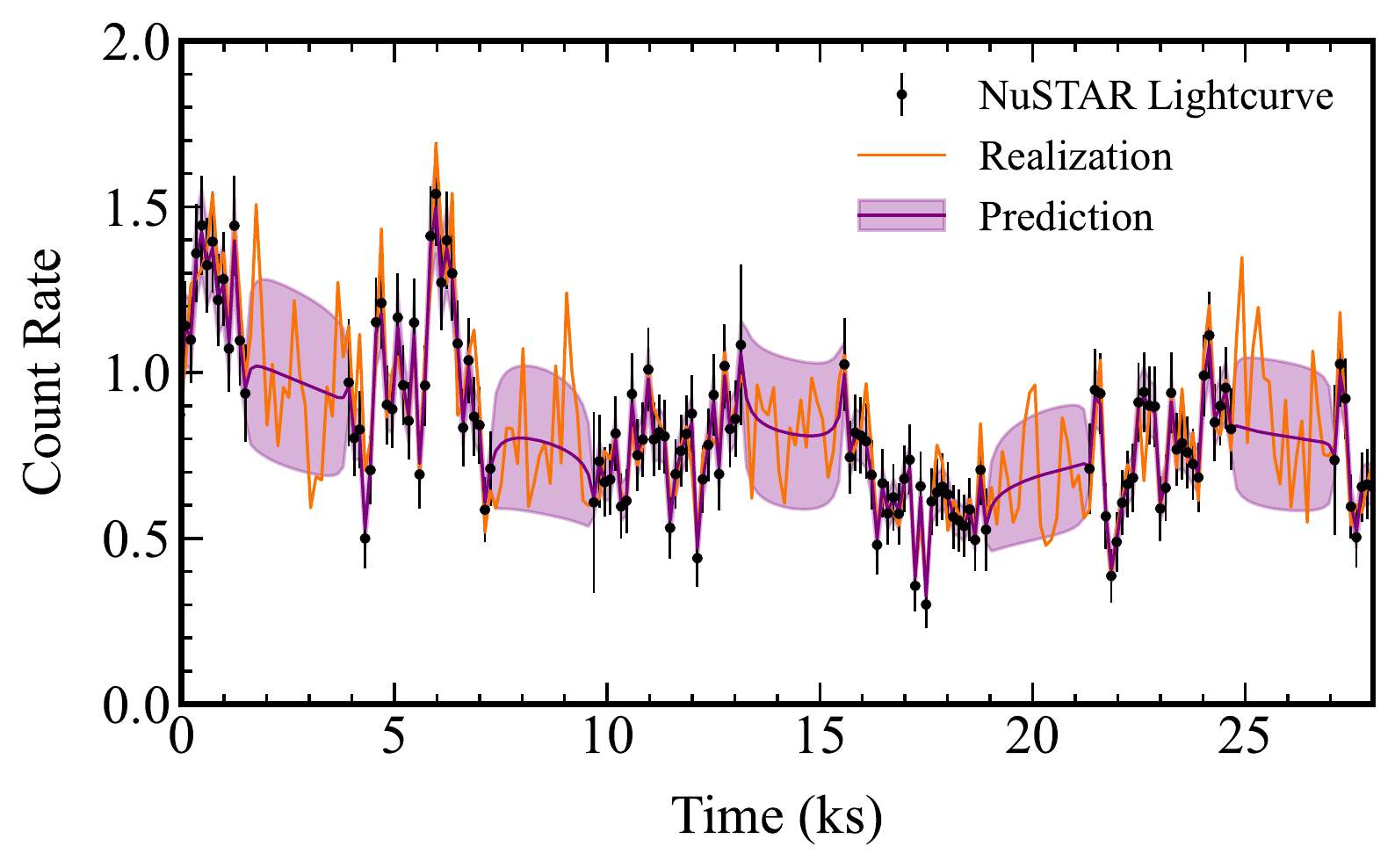}
    \caption{An example of a single lightcurve realization drawn from the trained GP posterior is shown in orange, with the prediction formed by averaging over 5000  samples shown in purple with a 1$\sigma$ shaded region in this distribution. In practice, we do not average over the samples themselves as shown here (we wish to use the variability in each sample) in order to calculate the time lags, but instead average the Fourier products across realizations. }
    \label{fig:gp_reals}
\end{figure}

We aim to analyze and model Ark~564's time lags, which we uncover using a Fourier approach that extracts the lags/leads of correlated variability between a lightcurve in each energy band and a common reference band. Resolving the time lags with a Fourier approach requires continuously sampled lightcurves (without data gaps). As such, we employ Gaussian processes to model the observed variability in our lightcurves in each energy band in order to draw continuous realizations of the data, or \textit{samples}, that include the data missing in the gaps. We implement Gaussian processes only on the NuSTAR data, as our XMM-Newton data is continuous over the timescales of interest, thus allowing for immediate use of standard Fourier-resolved timing analysis following \cite{Wilkins_2019}.

The lowest frequency one can probe with Fourier-resolved timing methods is set by the length of the longest continuous lightcurve segment. In the case of our NuSTAR data, the accessible frequency range is greatly restricted by data gaps occurring roughly every 90 minutes when the target is occulted by the Earth: $\sim10^{-4}$~Hz using the continuous segments versus $\sim 10^{-6}$~Hz if our longest NuSTAR observation were continuous. Gaussian processes allow us to access these lower frequencies of interest by drawing continuous lightcurve realizations that include data in the gaps. 

A \textit{Gaussian process} is formally defined as a collection of random variables, any finite number of which have a joint Gaussian distribution \citep{10.5555/1162254}.  In our case, these random variables are simply the count rate values of our time series, and the Gaussian process acts as a prior over a function of count rates $f(t)$ at time $t$. 

Following \cite{Wilkins_2019}, we have a data vector of count rates $\textbf{d}$ at times $\textbf{t}$ from NuSTAR. We assume this data is a realization of a Gaussian process, meaning we assume the vector $\textbf{d}$ has been drawn from a multivariate Gaussian distribution with mean function $m(t)=\mathbb{E}[f(t)]$ and the covariance between any two entries in the time vector is given by the covariance function, formally known as the \textit{kernel function}  $k(t,t’)=\mathbb{E}[(f(t) - m(t))(f(t^\prime) - m(t^\prime))]$. To generate other realizations of the data vector $\textbf{d}$, which we denote by the data vector $\textbf{d}_*$, at different times $\textbf{t}_*$, we first need to define a model for the mean function and one for the kernel function.

Our lightcurve data is used as the training set that informs these two functions, where ``training" generally describes the process of using our data to define these functions. As per standard practice, we first standardize the training set data by subtracting the mean from the lightcurves and dividing by the standard deviation. As such, we assume that $m(t)=0$. 

The kernel function describes how the data deviates from the mean function and is thus essential to an accurate description of observed variability. There are many functional forms for the kernel function, but we consider only those that are stationary, meaning they depend only on the time difference between points such that $k(t,t’) = k(t'-t)$ is a function of a single variable. Note that, in this stationary case, the kernel  $k(t'-t)$ is simply the auto-correlation function of the lightcurve. Each form of the kernel function has different \textit{hyperparameters} $\theta$ that encode a property of the variability, such as length scales and amplitudes. We determine the values of the kernel function's hyperparameters through a process known as \textit{hyperparameter optimization}.  This process consists of finding the set of hyperparameter values that maximizes the probability of the model given our training data, which is quantified by the marginal likelihood \citep{10.5555/1162254}. In practice, however, it is common to instead minimize the negative log marginal likelihood (NLML)
\begin{eqnarray}\label{eq: nlml}
-\log p(\mathbf{y}|\mathbf{t}, \mathbf{\theta}) &=& \frac{1}{2}\mathbf{y}^\top (K_\theta + \sigma_n^2I)^{-1}\mathbf{y} \\
&+& \frac{1}{2}\log |K_\theta + \sigma_n^2I| \nonumber \\ 
&+& \frac{n}{2}\log 2\pi \nonumber
\end{eqnarray}where $\textbf{y} = \textbf{d} - m(t)$ is the set of normalized observed count rate values and the entries of the covariance matrix $K_\theta$ are $K(i,j) = k( t_i , t_j | \theta )$, whose hyperparameters $\theta$ have been optimized using $n$ training points. The kernel function $k(t, t')$ generates the elements of the covariance matrix. $I$ is the identity matrix, and $\sigma_n^2$ is the expected contribution of noise to the variance of each data point. We note that treating the Poisson noise as Gaussian is valid at high count rates, which is not the case for our NuSTAR data above 14~keV. As a check, we rebinned the data to ensure $>$20 counts/bin before re-running the GPs. We find that the resulting time lags above 14~keV remain within 1$\sigma$ of those computed originally with the lower-count-rate time binning, which have larger uncertainties as well.

The first term in the NLML equation is the only term that features the training set data (our observed count rate values) and acts to motivate the model's fit to the data. The second term punishes overly complex kernel functions, and the third term is a normalization constant. Since the Poisson noise affecting each point is uncorrelated, we account for noise in the model by adding a diagonal term ($\sigma_n^2I$) to the covariance matrix. 

In order to generate realisations of the lightcurves with no gaps with data points $\textbf{d}_*$, we make random draws of the conditional distribution $(\textbf{d}_*|\textbf{d})$ from the multivariate Gaussian distribution defined by the optimised kernel function and the observed data vector \textbf{d} \citep[equation 5 from][]{Wilkins_2019}:
 \begin{eqnarray}\label{eq:cond_dist}
(\mathbf{x}|\mathbf{y}) \sim \mathcal{N}(\mu_x + K_{xy} K_{yy}^{-1}(\textbf{y} - \mu_y), K_{xx} - K_{xy} K_{yy}^{-1} K_{yx}) \nonumber
\end{eqnarray}where $\textbf{x} = \textbf{d}_*$ and $\textbf{y} = \textbf{d}$ and $\mu$ is the mean of these vectors. $K_{xy}$ denotes the covariances between elements of $\textbf{x}$ and $\textbf{y}$, which is calculated by evaluating the kernel function between our obsrved time bins $\textbf{t}$ and the time bins $\textbf{t}_*$ at which we draw lightcurve realizations.

Rather than training a separate Gaussian process on each individual lightcurve, we simultaneously fit the model kernel function to the entire data set in each energy band. This informs a single set of shared hyperparameters, effectively solving multiple machine-learning tasks (regression) with a single model---an approach known as \textit{multi-task learning}. Training using the entire data set allows the model to best learn the observed variability on both short and long timescales. The advantages and assumptions of this approach are discussed more in the following subsection.

After training the model, we draw 5000 evenly-sampled lightcurve samples from the Gaussian process posterior for each observation including points in the data gaps. An example of one of these samples is shown in Figure \ref{fig:gp_reals}. We find the resulting time lags converge when drawing this number of samples, meaning that drawing additional samples does not affect the resulting time lags. As discussed in Section \ref{sec:fourier}, we determine the time lags between energy bands by first averaging the Fourier products (cross-spectra and power-spectra) obtained from the lightcurve samples instead of averaging the time lags across the samples themselves. However, we find the results from these two approaches to be consistent.

This is the first implementation of multi-task learning for X-ray timing analysis, and we developed our architecture for this approach to model training and drawing from the posterior by combining and modifying the tools from \texttt{Scikit-learn}\footnote{\href{https://scikit-learn.org/}{https://scikit-learn.org/}} and the X-ray timing analysis package \texttt{pyLag}\footnote{\href{http://github.com/wilkinsdr/pylag}{http://github.com/wilkinsdr/pylag}} \citep{Wilkins_2019}. 

\subsection{Multi-task learning for hyperparameter optimization}
Previous applications of Gaussian processes for X-ray lightcurves modeled the variability of each observation independently \citep[e.g.][]{Wilkins_2019, Griffiths_2021}. Here, we train a shared set of hyperparameters by optimizing the summed NLML of all NuSTAR observations in each energy band of interest. This is equivalent to simultaneously fitting the model kernel function to the entire data set in each energy band. 

Multi-task learning has been shown to improve how effective the model is at predicting missing data, known as model generalization \citep{10.1023/A:1007379606734}. Determining a single set of hyperparameters by training on all of our observations is especially advantageous for shorter observations, such as our 40~ks NuSTAR observation from 2018, which has far fewer data points to train its own set of hyperparameters. It is important that the Gaussian process models variability across a wide range of timescales; however, training a set of hyperparameters with only a short observation biases against modeling longer-timescale variability \citep{10.1023/A:1007379606734}. 

Smaller training sets are also more likely to have multiple local optima similar in magnitude in the NLML function, corresponding to different and often unphysical interpretations of the data \citep{10.5555/1162254}. While we perform optimization multiple times in attempt to avoid such local optima, training using all observations largely mitigates this issue, resulting in local NLML optima now differing by orders of magnitude. Lastly, longer observations also have larger contributions to the total NLML and thus have a greater weight on the model as expected.

Our approach using multi-task learning assumes that our lightcurves can be characterized by a single description of the observed variability on timescales up to the length of the observation, or, equivalently, a single, shared kernel function. This would require an assumption that there are no fundamental differences across observations regarding the underlying variability processes and the characteristic timescales that the hyperparameters describe. As is common practice in X-ray timing analysis, we assume stationarity of Ark~564, which is motivated by the general consistency in Ark~564's power spectral density (PSD) across our observations. As a check, we compute the Lomb-Scargle periodogram of each NuSTAR observation in the frequency range of interest (0.01--0.9~mHz), and find that fitting each periodogram with a power law results in fit parameters that agree within $1\sigma$. We find similar consistency when computing the PSDs above 0.3 mHz using the continuous NuSTAR data between the gaps and fitting each with a power law. 

\subsection{Selecting the kernel function}\label{subsec:kernels}
Choosing a proper functional form for the kernel function is crucial to accurately modeling the observed variability, and has been found to affect the significance of lag detection \citep{Griffiths_2021}. We consider three common forms for the kernel function to model the variability of Ark~564: the squared exponential (SE), rational quadratic (RQ), and Mat\'ern kernel. We refer the reader to \cite{Wilkins_2019} for an introduction and details on these kernel function forms.

We minimize the NLML function to determine the set of hyperparameter values given our lightcurves. This process is performed separately in each energy range of interest and for the broader reference band. As such, we are interested in comparing how accurately each kernel form can model the variability in both the individual primary energy bands of interest and in the reference band. This is commonly done by comparing the probability that the model describes the training data, i.e. the minimized NLML values for the best-fit kernel. As such, we average the minimized NLML values for the primary energy bands of interest, as well as for the reference band, shown in Table \ref{tab:nlml}.

While we do find the choice of kernel form to impact lag detection, the effects of this choice are much smaller in our case than those found by \cite{Griffiths_2021}. This is likely due to the difference in data sampling/length of their Swift monitoring campaign versus our NuSTAR data. The uncertainties on the lowest-frequency lags are roughly constant across kernel forms (within roughly 20\% on-average and 10\% for the high-frequency lags). The amplitude of the lag is even more consistent across kernel forms (within 15\% for the lowest-frequency lags and 10\% for the high-frequency lags). In summary, the time-lag results from using any of the investigated kernel forms are generally consistent in all three frequency ranges of interest.

We conclude that the RQ kernel best captures the variability of our lightcurves, given that this kernel results in the lowest optimized total NLML value, with the Mat\'ern-$\frac{1}{2}$ kernel inching behind. \citet{Wilkins_2019} similarly finds the RQ kernel to be the most successful for lag recovery when simulating \textit{NICER} lightcurves of Ark~564 with a 200s lag and low-Earth orbit data gaps. The author concludes that the accuracy achievable with the RQ kernel is sufficient for analysis of data available from current missions. Our NLML results are also generally consistent with those from \citet{Griffiths_2021}, who found the RQ and Mat\'ern-$\frac{1}{2}$ kernel functions to yield the most well-specified models for coarsely sampled, simulated X-ray lightcurves of Mrk 335. 

\begin{deluxetable}{lcc}[t!]
\tablecaption{Optimized NLML values averaged across all primary energy bands of interest, as well as those averaged across the broader reference band lightcurves for each kernel form. Lower NLML corresponds to higher probability of the model given the training data. \label{tab:nlml}}
\tablehead{\colhead{Kernel Form} &
\colhead{NLML (Primary)} & \colhead{NLML (Reference)}
}
\startdata
SE & 7182.4 & 5347.7 \\
Mat\'ern-$\frac{1}{2}$ & 7059.3 & 5233.5 \\
Mat\'ern-$\frac{3}{2}$ & 7090.3 & 5518.8  \\
RQ & \textbf{7008.5} & \textbf{5202.0} \\
\enddata
\end{deluxetable}

\section{Fourier-resolved Timing with Gaussian processes} \label{sec:fourier}

After training the Gaussian process and drawing continuous lightcurve samples, we apply standard Fourier-resolved timing analysis, described further by \citet{Uttley_2014}, to determine the time delays between correlated variability in different energy bands with respect to a common reference band. 

The lowest frequency one can probe is limited by the length of the longest continuous lightcurve segment (the length of the observation in an ideal case). The highest frequency attainable is set by the Nyquist frequency $f=1/2\Delta t$ for sampling rate $\Delta t$, but dominating Poisson noise greatly lowers this limit in practice. 

For two lightcurves $x(t), y(t)$ with corresponding Fourier transforms $\tilde{X}(f)=|\tilde{X}(f)|e^{i\phi_X}$ and $\tilde{Y}(f)=|\tilde{Y}(f)|e^{i\phi_Y}$, the cross-spectrum $\tilde{C}(f)$ is given by 
\begin{equation}
    \tilde{C}(f) = \tilde{Y}^*\tilde{X} = |\tilde{X}(f)||\tilde{Y}(f)|e^{i(\phi_X-\phi_Y)}
\end{equation} 

We convert the argument of the cross-spectrum (the phase lag as a function of frequency) to a time lag by dividing by the frequency, resulting in the \textit{lag-frequency spectrum}
\begin{equation}
    \tau(f) = \frac{\arg[\tilde{C}(f)]}{2\pi f}
\end{equation}

The lag-frequency spectrum allows us to probe correlated variability between two lightcurves as a function of timescale. A positive time lag at frequency $f$ indicates (by convention) that the variability process with corresponding timescale $1/f$ is first observed in lightcurve $X$ and is seen lagging in lightcurve $Y$. We refer to lightcurve $X$ as the primary energy band lightcurve, and lightcurve $Y$ to be the reference band lightcurve. 

We create the cross-spectrum between each primary energy band of interest and a common reference before averaging over a broader frequency bin to optimize the signal-to-noise, resulting in the \textit{lag-energy spectrum}. We emphasize that we are interested in the relative lag between energy bins to gain information on correlated variability in different energy bands with respect to a common reference band, rather than the amplitude of the lags themselves.

\begin{figure*}[ht!]
    \centering
    \includegraphics[width=0.68\textwidth]{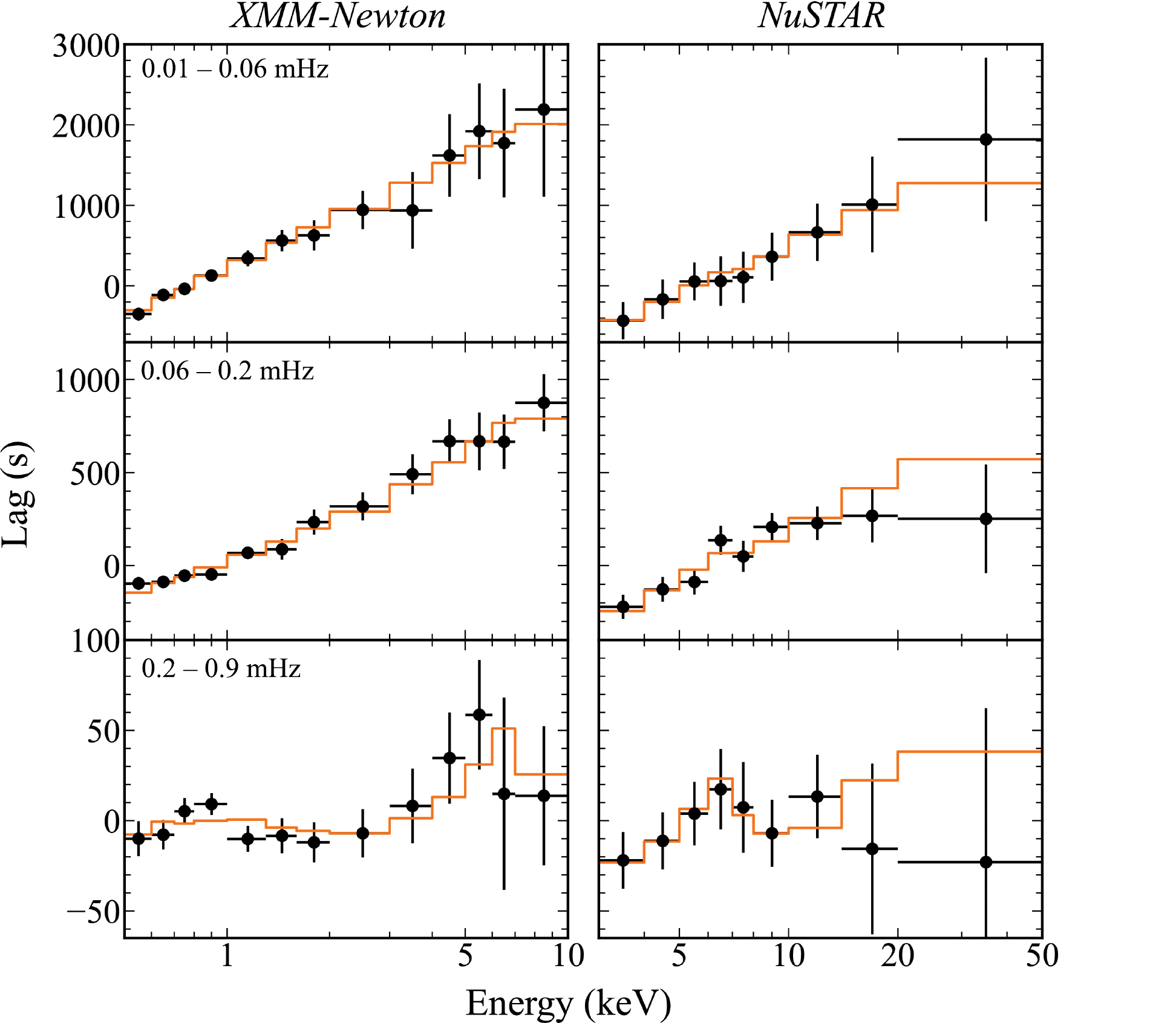}
    \caption{Lag-energy spectra from each instrument in three frequency ranges (black), modeled with \texttt{reltransDCp} by fitting the flux spectra with the lag-energy spectra (orange). The XMM-Newton time lags are calculated using the entire 0.5--10~keV energy range, versus 3--14~keV for the NuSTAR time lags to maximize signal-to-noise.}
    \label{fig:lags_fit}
\end{figure*}
The reference band is taken to be the broadest possible reference band (across all energies with good signal-to-noise) to minimize uncertainty associated with Poisson noise. We also subtract the primary energy band lightcurve from the reference band lightcurve to remove Poisson noise in the primary band that would be correlated with itself in the reference band. For XMM-Newton observations, we use the entire 0.5--10~keV energy range for the reference band, and 3--14~keV as the reference band for our NuSTAR lightcurves where signal-to-noise is greatest.

In the case of our continuous XMM-Newton lightcurves, we compute these timing products immediately, without Gaussian processes. For each data-gap-ridden NuSTAR observation, we draw 5000 samples of both the primary energy band and the primary-band-subtracted reference band from the Gaussian process posterior. We calculate the cross-spectrum and power spectra from each sample pair before averaging the spectra across the 5000 sample pairs, resulting in three final spectra required to calculate the time lag and associated uncertainty. Uncertainty on the lags is calculated from coherence as outlined by \citet{Uttley_2014}.

The lag-energy spectrum reveals how different energy bands contribute to correlated variability on different timescales, and is thus an important tool for studying the causal relationship between different spectral components \citep{Uttley_2014}. We thus combine the Fourier products given that we have assumed stationarity of the source on the timescales spanned by our observations, motivated by the cross-spectra and power spectra being consistent across observations. In each energy bin, we take all points in each observation's cross-spectra and pair of power spectra (the power spectra are used to calculate coherence and thus uncertainty on the lags) that fall in a frequency range of interest, average, and convert to a final time lag. This is particularly valuable for our NuSTAR observations to reduce the larger time-lag uncertainties imposed by Gaussian processes (see Section \ref{sec:fourier}). Since there is no prior placed on correlations between energy bands, sampled data in the gaps will be uncorrelated with that in another energy band, resulting in lower coherence and thus larger errors. 
Ark 564 is one of the only sources with a low enough mass that allows us to perform the test of comparing the reverberation lags above $\sim 0.03$ mHz computed using GPs versus the immediately available continuous NuSTAR segments. The results from both approaches are consistent, as shown in Figure \ref{fig:gp_vs_chunk}, but using Gaussian processes results in larger errors by roughly 40\% on average below 10 keV.

We perform simulations similar to those of \citet{Wilkins_2019} to evaluate the effects of the Gaussian processes on time lag recovery in the case of our specific observations. We simulate lightcurves with lengths, means, and standard deviations matching our observations using the method of \cite{timmer}. The ``reverberation'' lightcurves are generated by convolving the original lightcurves with a $\delta$-function, shifting the lightcurves by 100s (on the order of the observed Fe K lag). We introduce data gaps replicating those from the NuSTAR low-Earth orbit and generate Gaussian process realizations including data in the gaps. We find results consistent with \cite{Wilkins_2019}: we are able to recover the 100s lag within 5\% at all frequencies in which reverberation is probed in this paper ($0.06-0.9$~mHz). 

\section{Results} \label{sec:results}
\begin{figure*}[ht!]
    \centering
    \includegraphics[width=\linewidth]{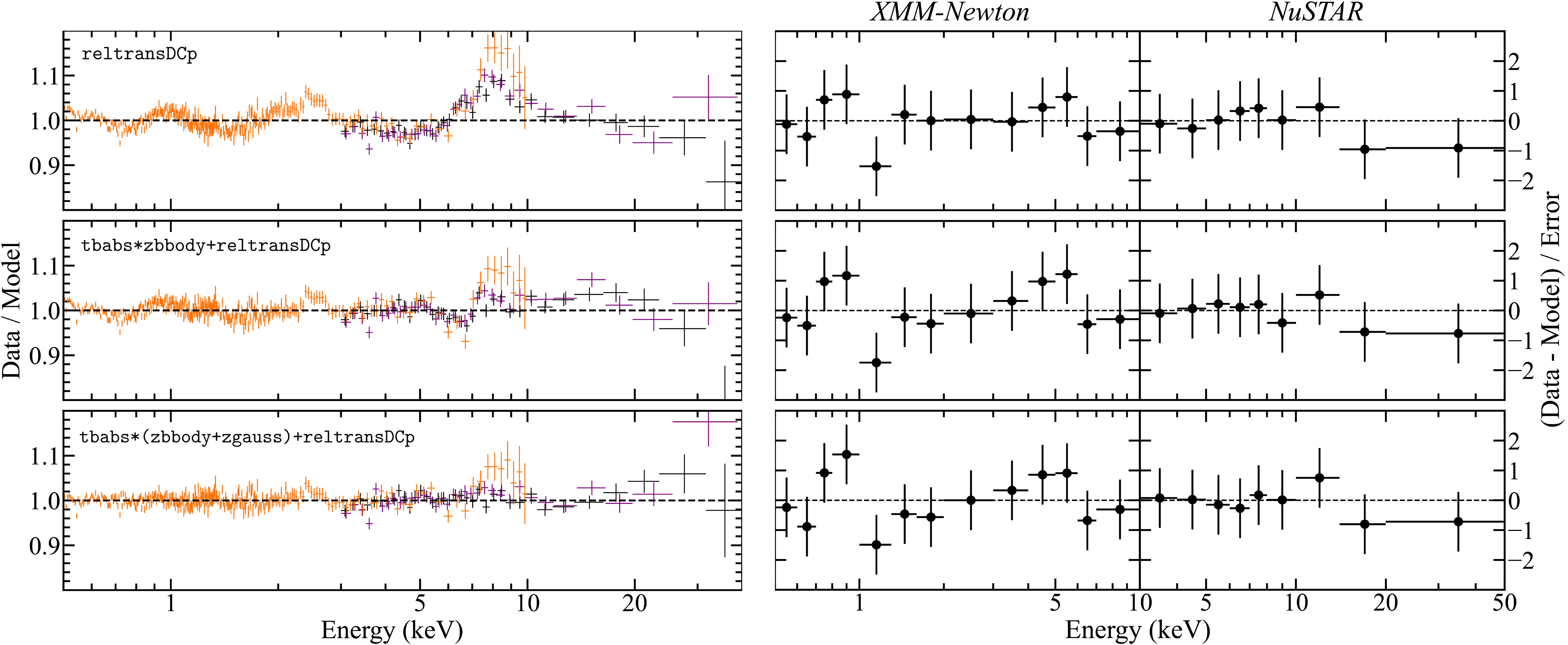}
    \caption{(Left) Data-to-model ratios of the XMM-Newton (orange) and NuSTAR FPMA, FPMB (black, purple) flux spectra combined across observations when fit simultaneously with the time lags for each of our three models. (Right) The resulting fit residuals on the high-frequency time lags (0.2--0.9~mHz) for each instrument is shown in the right six panels. The inclusion of a single-temperature blackbody (middle) with \texttt{zbbody} dramatically improves the fit, but show residuals on a feature near 1~keV, motivating the inclusion of an additional Gaussian component (bottom), further improving the residuals on the reflection features. The model of the time lags does not change significantly, but struggles to fully replicate the feature near 1~keV.}
    \label{fig:3models}
\end{figure*}

The lowest-frequency lag-energy spectra (the top row of spectra in Figure \ref{fig:lags_fit}) exhibit the canonical hard lags, which monotonically increase with energy. The intermediate frequency range (the middle row of spectra in Figure \ref{fig:lags_fit}) similarly shows the hard lags, but begins to show faint signs of the high-frequency reverberation lags, namely a potential Fe K lag in the 6--7~keV NuSTAR bin. Interestingly, in the intermediate frequency range from 0.06--0.2~mHz, we see that the hard lags change in slope and level out at 10~keV, which can be seen only in the NuSTAR band. This atypical change in slope in the hard lags could be related to the appearance of the Compton hump in the spectra, or due to the particularly low electron temperature of the corona (10--20~keV). 

The high-frequency spectra (the bottom row of spectra in Figure \ref{fig:lags_fit}) show tentative evidence for the soft lag features associated with reverberation. A previous 500~ks XMM-Newton campaign of Ark~564 showed significant detections of a soft lag \citep{2012ApJ...760...73L} and an Fe K lag at 99.94\% significance \citep{2013MNRAS.434.1129K}, but in our shorter campaign, we do not make a significant detection of the iron~K line. That said, the lags do increase from 4--10~keV and peak at 6--7 keV, consistent with previous results. We also see a broad feature centered at 0.8--0.9~keV in the high-frequency lags, which is consistent with a feature that we observe in the flux spectrum. The potential origin of the feature is discussed in Section \ref{sec:discussion}.

While in practice, we will use the spectrum and the Fourier-resolved time lags to constrain model parameters, here we focus on the significance of reverberation in just the high-frequency lags. We perform a simple test for the statistical significance of the reverberation features by considering the null hypothesis that the observed lag spectra in the highest frequency range are comprised of only the continuum, modeled in \textsc{reltrans} as a pivoting power-law (i.e. the boost parameter is set to 0 so no reverberation is included). For this test, we assume values for the parameters found from fitting the flux spectra alone. We compare the resulting best-fit ($\chi^2_\nu/\nu = 1.36/10$, where $\chi^2_\nu$ denotes the reduced chi-squared for $\nu$ degrees of freedom) to that found when including non-zero boost with the pivoting power-law. The latter model (i.e. including reverberation) results in a $\Delta \chi^2=-3.89$ for one additional degree of freedom ($\chi^2_\nu/\nu = 1.01/9$), corresponding to a significance of 91.1\% using an F-test.

The 0.2-mHz lower bound of the range of frequencies considered for reverberation (i.e. the highest frequency range) was selected due to limited NuSTAR statistics from employing GPs.  Contributions from the hard lag at this lower frequency appear to be flattening the lag-energy spectrum below 2 keV. When we instead isolate higher frequencies, we find prominent soft lags from the soft excess consistent with those that have been seen previously in this source \citep{2013MNRAS.434.1129K}.  

We model the XMM-Newton and NuSTAR flux spectra and lag-energy spectra simultaneously using the relativistic reverberation mapping model \textsc{reltrans} \citep{reltrans}. We fit six lag-energy spectra in total with the flux spectra: one lag-energy spectrum per observatory in each of three contiguous frequency ranges spanning the accessible frequency range (0.01--0.9~mHz). The low-frequency hard lags are modeled by a pivoting power-law component discussed by \cite{2018MNRAS.475.4027M, 2021MNRAS.507...55M}, which includes 3 additional free parameters per frequency range: $\phi_{A}$ (a normalization phase), $\phi_{AB}$ (phase difference between the spectral index and the normalisation variability), and $\gamma$ (amplitude ratio between the spectral index and the normalisation variability). $\phi_{AB}$ and $\gamma$ are tied between lag-energy spectra in the same frequency range, whereas the normalization phase $\phi_{A}$ is left untied. 

We fit the flux and lag spectra (by minimizing the summed $\chi^2$) with \texttt{reltransDCp}, a flavor of \textsc{reltrans} that implements a physically motivated thermal Comptonization continuum using \texttt{nthcomp} \citep{1996MNRAS.283..193Z, 1999MNRAS.309..561Z} to model the coronal emission with the parameters power-law photon index $\Gamma$ and electron temperature $kT_e$. The reflection in the model originates from the \texttt{XILLVER} grids \citep{2013ApJ...768..146G}. The model calculates the energy shifts and light-crossing delays, including general relativistic effects, of the X-rays emitted by a lamppost corona positioned on the spin axis of the central black hole. The X-ray reflection spectrum is computed using the most updated rest-frame reflection emission convolved with the general relativistic ray-tracing transfer function computed from the lamppost corona. The model correctly accounts for the instrument response from both instruments when computing the lag-energy spectrum. This version is also capable of probing the disk's electron density up to $\log(n_e/\text{cm}^{-3})=20$. \texttt{reltransDCp} is not yet publicly available, but will be included in public releases in the future. The model is capable of fitting the time lags and the flux spectra by self-consistently modeling the direct coronal continuum and the corresponding reflection spectrum. We probe model consistency for describing the time lags and the flux spectra with a single model for the accretion disk by fitting these data products simultaneously, which was fruitful for \cite{2020MNRAS.498.4971M}. We present the resulting constraints on properties of the accretion disk and black hole, including a mass estimate from fitting the time lags. The fit parameters and model consistency are discussed further in Section \ref{sec:discussion}.

Galactic absorption is accounted for in the blackbody component only, since it is already accounted for self-consistently within \texttt{reltransDCp}. The frozen fit parameters are the cosmological redshift (0.02468), the inner and outer radii of the disk (at the innermost stable circular orbit (ISCO) and $1000 R_g$, respectively), and a maximum value for the spin, which is consistent with previously reported values \citep{10.1093/mnras/sts227, 2018PhRvD..98b3018T, density}. The ionization of the disk ($\log(\xi)$) is assumed to be constant. 

Our initial best fit with \texttt{reltransDCp} alone gives a poor description ($\chi^2_\nu/\nu= 1.48/3068$) of the flux spectrum, especially the Fe K feature. The poor quality of the fit is shown by the data-to-model ratio shown in the top panel of Figure \ref{fig:3models}. The fit improves drastically when excluding energies below 2~keV and refitting ($\chi^2_\nu/\nu=1.01/2763$).

In order to model the soft excess below 2~keV, we include a phenomenological single-temperature blackbody component ($kT = 0.11$~keV) to the model using \texttt{zbbody}, which contributes to the flux spectra but not the time lags. We account for galactic absorption \citep[$n_{\text{H,Gal}} = 5.3 \times 10^{20}$ cm$^{-2}$ ;][]{2005A&A...440..775K} with \texttt{tbabs}, using abundances from \cite{2000ApJ...542..914W}. This model (\texttt{tbabs*zbbody+reltransDCp} in {\sc xspec} syntax) achieves a much more successful description of the flux spectra in particular when including the soft excess below 2~keV ($\chi^2_\nu/\nu = 1.21/3065$). All fit parameters except the source height of the lamppost corona, inclination, ionization, and boost remain within error with the inclusion of the blackbody. The boost factor adjusts the reflection fraction resulting from the assumed point-like lamppost corona, with values $>1$ causing a stronger reflection component than expected in the lamppost geometry and $<1$ returning a weaker-than-expected reflection \citep{reltrans, 2020MNRAS.498.4971M}. The model is driven to a higher source height (from $2.1^{+0.7}_{-0.3}  R_g$ to $10.3^{+0.7}_{-0.7} R_g$), higher inclination (from $36.7^{+0.9}_{-0.8}$ to  $44.8^{+0.7}_{-0.7}$),  lower ionization ($\log(\xi)$ from $3.30^{+0.01}_{-0.01}$ to $3.16^{+0.01}_{-0.01}$, where $\xi$ has units of erg cm/s), lower boost (from $0.61^{+0.01}_{-0.02}$ to $0.49^{+0.01}_{-0.01}$). These changes, in addition to a near-maximal iron abundance ($\sim$10 times solar iron abundance) in both cases, are likely attempting to fit the aforementioned feature just below 1~keV in both the flux spectra and the time lags. Nonetheless, the new fit leaves residuals in the flux spectra throughout the soft excess, especially around the $\sim$1 keV feature, as shown in the middle panel in Figure \ref{fig:3models}. The model continues to underestimate the Fe K$\alpha$ line, even with maximal iron abundance.

\begin{figure*}[ht!]
    \centering
    \includegraphics[width=0.64\textwidth]{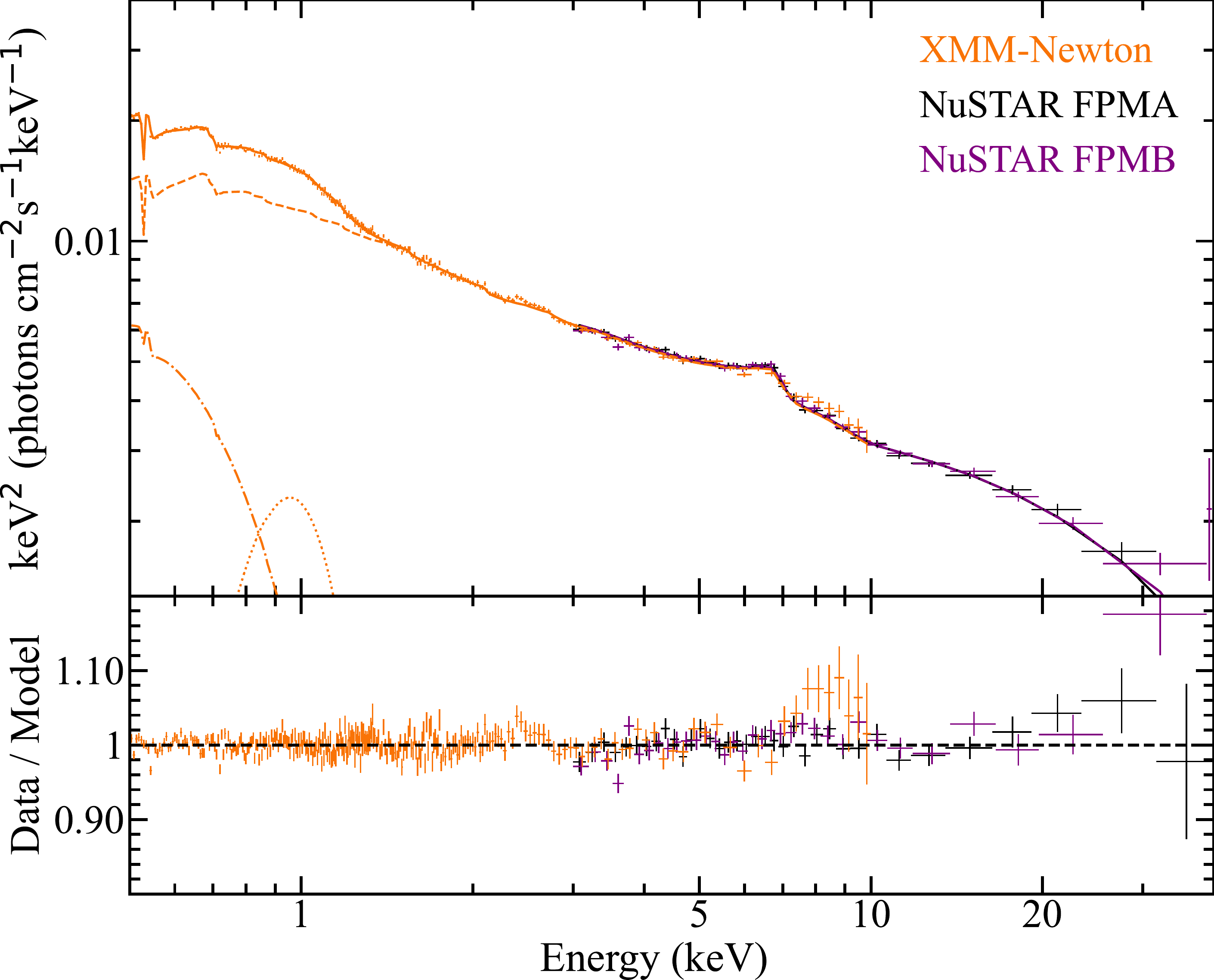}
    \caption{The XMM-Newton and NuSTAR flux spectra from 0.5 to 40~keV with the best-fit model (\textit{upper}) and corresponding data-to-model ratio (\textit{lower}). The total model \texttt{(tbabs*(zbbody+zgauss)+reltransDCp)} is shown by a solid line, with model components \texttt{tbabs*zbbody} (blackbody component to fit the soft excess), \texttt{tbabs*zgauss}, and \texttt{reltransDCp} shown by dash-dotted, dotted, and dashed lines, respectively.}
    \label{fig:spectra_fit}
\end{figure*}

Because we were unable to find an acceptable fit with just a blackbody soft excess component, we include a phenomenological Gaussian component using \texttt{zgauss} to assist the model with the $\sim$1~keV feature in the flux spectra, leaving the center and width of the Gaussian as free parameters. The feature is not a narrow line, and, likewise, the Gaussian is broad ($\sigma = 0.18$) and is centered at 0.9~keV. The model (\texttt{tbabs*(zbbody+zgauss)+reltransDCp}) includes the blackbody and again accounts for galactic absorption on the Gaussian and blackbody components. Adding this Gaussian component dramatically improves the residuals around the reflection features, especially the soft excess, in the flux spectra. The fit on the time lags shows minor improvements, particularly in the intermediate frequency range. The resulting best fit of this final model is shown in Figure \ref{fig:spectra_fit}, with a fit statistic of $\chi^2_\nu/\nu = 1.08/3062$. The fit is dominated by the flux spectra, with the flux and the time lags contributing a $\chi^2$ of 3265.13 (in 3026 bins) and 26.98 (in 66 bins), respectively. The best-fit parameters are comparable to those when fitting without the Gaussian, with the exception of iron abundance. The component removes the need for the reflection to replicate the $\sim$1~keV feature with near-maximal iron abundance, lowering the iron abundance from 10 to 2.25 times solar iron abundance. We find that all of the best-fit parameters found when modeling the flux spectra above 2~keV with \textsc{reltrans} (without additional components) lie within $2\sigma$ of those found when fitting 0.5--40~keV with the additional blackbody and Gaussian components, with most values consistent within $1\sigma$. Table \ref{tab:comp_table} in the appendix shows the fit parameters in these two cases.

While the Gaussian is phenomenological, similar features have been seen in other high-Eddington accretors, such as the NLS1 1H 1934-063 \citep{2022arXiv220406075X} and the extreme changing-look AGN 1ES 1927+654 \citep{2022arXiv220605140M}. We discuss further the potential physical origin of the feature near 1~keV in Section \ref{sec:discussion}.
 
 Warm absorbers have been seen previously in this source \citep{2004ApJ...603..456M, 2015A&A...577A...8G}. When we fit the RGS spectra from 0.4--1.5~keV with a blackbody and powerlaw, we see narrow absorption that is particularly strong around O VII and O VIII. Further modeling of the RGS data of the warm absorbers is out of the scope of this paper, but we do see a broad excess from 0.8 to 1.5 keV, which is also seen in the EPIC-pn data. Additional future work is warranted on modeling these features. 

The model is successful at reproducing both the hard and soft lags across the entire frequency range, except for residuals on the feature near 0.9~keV in the high-frequency XMM-Newton spectrum. The lag spectra are generally consistent with those computed from the XMM-Newton archival data, although minor changes in the high-frequency lags below 1~keV render the soft-excess lags to be less prominent than those found previously \citep{2013MNRAS.434.1129K}. Changes in the lag-energy spectra have been observed on month-long timescales in this source \citep[see Figure 5 in ][]{2013MNRAS.434.1129K} and in other sources \citep[e.g.][]{2020NatAs...4..597A}. Proposed causes include changes in the geometry of the corona, which we are unable to distinguish in our case: if the best-fit model from this paper is applied to the archival lags, a similar fit statistic is obtained (the reduced chi-squared increases by less than 0.2), with residuals on the high-frequency lags located below 1~keV in both cases. 

 \begin{figure*}[t!]
    \centering
    \includegraphics[width=0.7\linewidth]{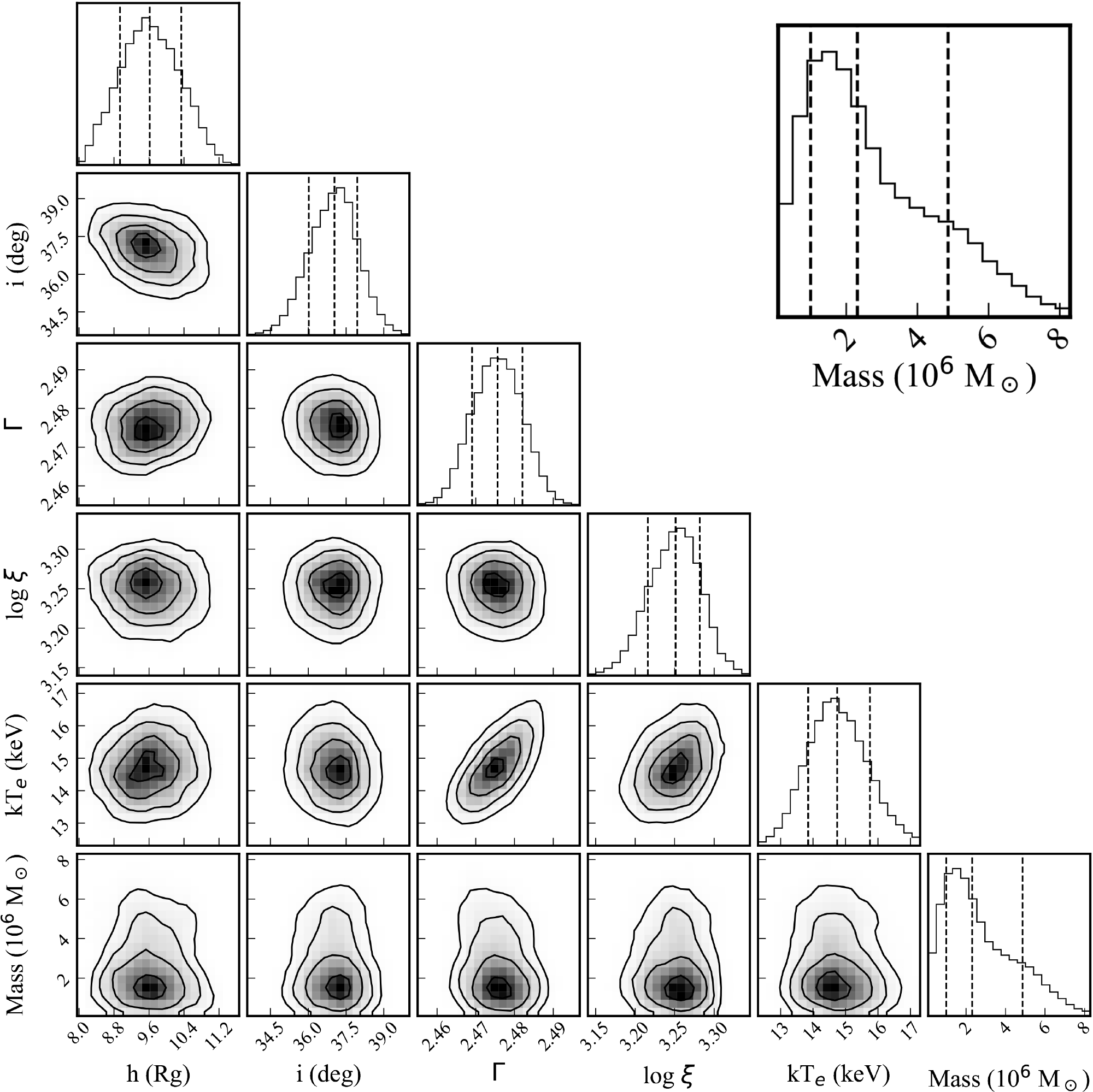}
    \caption{Posterior distributions from the MCMC of fitting the time lags with the flux spectra. The plots on the diagonal show the histogram for each model parameter, dashed vertical lines corresponding to the median and $1\sigma$ from the median. The off-diagonal plots are projections of the 2D distributions between any two parameters, with contours of equal probability. An enlarged 1D mass posterior is shown in the upper right.}
    \label{fig:mcmc}
\end{figure*}

The reported parameter values and their associated uncertainties are determined using a Monte Carlo Markov Chain (MCMC) analysis, using the Goodman-Weare algorithm in \texttt{xspec}. We ran three chains, each with a different initial distribution for the walkers to ensure that the walkers are not trapped in local extrema. Each chain consists of 200 walkers in-total with 15,000 iterations each, with the walkers initially either distributed 1) normally with mean and covariance information from the best-fit, 2) normally with the same normal distribution but with the covariance rescaled by 0.2, or 3) uniformly about the best-fit parameters. For the chain with the first aforementioned initial distribution, we burn the first 25\% of the iterations per walker, whereas the other two chains converge much faster and require a burn-in of only 10\% of iterations. It is after this burn-in that we find each MCMC has converged based on a rapidly decaying autocorrelation function and the Geweke convergence measure for every free parameter. We also find the posterior distributions from the first 10\% and last 50\% of the chains (excluding the burn-in period) are consistent, further indicating that the chain has converged in the first 10\% of iterations. We lastly combine the chains, finding that the three chains converge to the same stationary distribution based on the Gelman-Rubin statistic for each parameter across the three chains. The median of the resulting posterior distributions and associated $1\sigma$-errors are shown in Table \ref{tab:fit-par}. The posterior distributions, made using module \textsc{corner} \citep{corner}, for several fit parameters are shown in Figure \ref{fig:mcmc}.
\begin{deluxetable}{lc}[ht!]
\tablecaption{Median and 1$\sigma$-uncertainties on the fit parameters for fitting the flux spectra and lag-energy spectra simultaneously, determined from the MCMC posterior distributions. \label{tab:fit-par}}
\tablewidth{0pt}
\tablehead{\colhead{Parameter} & Time-avg. + Time Lags}
\startdata
$h$ $[R_g]$  & $9.6^{+0.7}_{-0.7}$ \\
$i$ $[\text{deg}]$ & $37.0^{+0.9}_{-1.0}$ \\
$\Gamma $ & $2.48^{+0.01}_{-0.01}$ \\
$\log(\xi$ $[\text{erg cm/s}])$& $3.25^{+0.03}_{-0.04}$\\
$A_{\text{Fe}}/A_{\text{Fe}_\odot}$ & $2.3^{+0.5}_{-0.4}$\\
$\log(n_e$ [cm$^{-3}$]) & $15.1^{+0.2}_{-0.1}$\\
$1/\mathcal{B}$ (Boost) & $0.50^{+0.03}_{-0.03}$\\
$kT_e$ $[\text{keV}]$  & $14.7^{+1.0}_{-0.9}$\\
$\text{Mass}$ $ [10^6 \text{M}_\odot] $ & $2.3^{+2.6}_{-1.3}$ \\
$\chi^2_\nu/\nu$  & 1.08/3062
\enddata
\end{deluxetable}

The best-fit source height ($9.6^{+0.7}_{-0.7}$) is typical for a $10^6-10^7 \text{M}_\odot$ black hole \citep{ 2016MNRAS.460.3076C, Kara_2017, 2020NatAs...4..597A, 2020MNRAS.498.4971M}. We find the temperature for the corona to be low at $14.7^{+1.0}_{-0.9}$~keV, consistent with the low coronal temperature reported by \cite{Kara_2017}. We find a relatively low electron density for the disk ($\log(n_e/\text{cm}^{-3}) = 15.1^{+0.2}_{-0.1}$). We were unable to find a better fit when requiring a higher value for the density, such as the density value for Ark 564 reported by \cite{density} ($\log(n_e/\text{cm}^{-3}) = 18.55$), or by using a steppar over density.

Modeling the time lags allows one to also estimate the mass of the black hole, from which we find the mass of Ark~564 to be $2.3^{+2.6}_{-1.3}\times 10^6 M_\odot$. This constraint is consistent with \cite{2009ApJ...702.1353D} ($1.7 \times 10^6 M_\odot$), \cite{2004AJ....127.3168B} ($2.6 \times 10^6 M_\odot$), and the upper limit found from optical reverberation mapping by \cite{2001ApJ...561..162S}. Our estimate is also within error from the upper mass limit reported by \cite{2009MNRAS.394.2141N}, which was determined from applying NLS1-corrections to mass estimates found using the scaling relation by \cite{2001A&A...377...52W, 2003ApJ...591..733B}. By applying a bolometric correction of 20 from \cite{2004MNRAS.351..169M} to our observed 2--10~keV luminosity ($2.40\times 10^{43}$ erg/s), this mass constraint corresponds to a bolometric luminosity of $L/L_{Edd} = 0.76-3.69$. As a result of fitting the lag-energy spectra in multiple frequency ranges, we do not see the degeneracy between black hole mass and corona height that has been observed when fitting lag-energy spectra in a single frequency range \citep{2014MNRAS.438.2980C, 2019MNRAS.488..324I}.

\section{Discussion} \label{sec:discussion}
We combined the flux and lag-energy spectra of our two new 115~ks XMM-Newton observations, as well as our new 410~ks NuSTAR observation with archival data, totalling 660~ks with NuSTAR and 230~ks with XMM-Newton. We apply the relativistic reverberation model \textsc{reltrans} to simultaneously model the flux spectra and six lag-energy spectra (one per observatory in each of three frequency ranges). The flux spectra required an additional blackbody and a Gaussian component in order to obtain a successful description of the soft excess, including an unknown feature at $\sim$1~keV, which also appears in the time lags. Including additional components to describe the flux spectra that are not included in the time lags allows us to better understand how features in the flux spectra manifest in the time lags, unlike previous works that found notable tension between the reflection features in the flux spectra and lag-energy spectra \citep[e.g.][]{2020ApJ...893...97Z, 2021ApJ...910L...3W}. 

\subsection{Fitting the soft excess}

In Section \ref{sec:results}, we were unable to adequately fit the flux spectra, especially the soft excess, with \textsc{reltrans} alone. The fit improves significantly by adding a blackbody component, which could be accounting for additional Comptonization off a warm corona \citep{2019ApJ...871...88G}.  We continue to see residuals around 1~keV in both the flux spectra and the time lags. The final fit of the Ark~564 flux spectra by \cite{2016MNRAS.460.3076C} shows similar residuals from this $\sim$1~keV feature. We found that including a broad Gaussian component to assist the model with this feature improved the fit residuals across the entire energy range and lowered the iron abundance from 10 to 2.25 times solar iron abundance. The best-fit parameters found from modeling the flux spectra above 2~keV with only \textsc{reltrans} are consistent with those found when fitting the entire spectrum with these additional components, with most values consistent within $1\sigma$. Here we discuss possible physical interpretations for why a broad Gaussian centered near 1~keV allows us to fit the soft excess. 

Super-solar iron abundances have been commonly found when fitting the reflection features \citep[e.g.][]{10.1111/j.1365-2966.2011.20356.x, garcia_2015, 2018cosp...42E1167G, 2020MNRAS.498.4971M}. Nonetheless, it has been suggested that such high iron abundance compensates for physics missing from the model in some cases \citep{2018cosp...42E1167G}. For instance, \cite{tomsick} found that their inferred super-solar iron abundance was attempting to account for high-density effects when modeling the spectra of Cyg~X-1. The broad $\sim$1~keV residual could be related to the model unable to fit the smeared Fe L reflection feature, which peaks near 1~keV \citep{2009Natur.459..540F}, and would explain why the feature arises in the high-frequency time lags. This could be related to modeling reflection off a razor-thin disk, a geometry less apt for describing high-Eddington accretion flows, where radiation pressure is expected to increase the scale height of the disk \citep[e.g.][]{1982ApJ...253..873B, 2011arXiv1108.0396S, 2014ApJ...796..106J, 2014MNRAS.439..503S, 2018ApJ...855..120T, 2019ApJ...884L..37L}.

Unidentified broad features around 1~keV have been seen in CCD spectra of several high-Eddington accretors. For instance, the 1~keV feature looks very similar to that detected in the NLS1 AGN 1H 1934-063 by \cite{2022arXiv220406075X}, the extreme changing-look AGN 1ES 1927+654 by \cite{2022arXiv220605140M}, as well as several ultraluminous x-ray sources (ULXs), especially those in the supersoft ultraluminous (SSUL) sub-class of ULXs \citep[e.g. NGC 247, NGC 6946, M1 and M101:][]{2016MNRAS.456.1859U, 2017MNRAS.467.2690E, 2021MNRAS.505.5058P}. Proposed solutions in these cases have included blueshifted O VIII ionization edges, or high-ionization Fe L or Ne IX-X absorption lines with velocities of $\sim$0.1--0.2c indicative of an optically thick wind \citep{2021MNRAS.505.5058P}. Indeed, previous studies of Chandra gratings observations of Ark~564 show blueshifted absorption lines consistent with a relativistic ($\sim$0.1c) outflow \citep{Gupta:2013in}.

It is possible that the residuals on the feature without the Gaussian is due to us not modeling the outflows, explaining why reflection alone fails to describe the feature and worsens the fit of the other reflection features in attempt to do so. This could also explain why we do not see the time lags being affected by including these phenomenological components, suggesting these additional components do not contribute to the variability on the timescales probed here. Nonetheless, additional future work is required, which includes modeling of the warm absorbers/outflows. 

\subsection{Consistency between time lags, time-avg. spectra}

 We include two additional components to assist in describing the flux spectra only. We are unable to meaningfully add these components to the time-lag fit, as components to the time lags contribute non-linearly. While this breaks the self-consistency of the simultaneous fit, we find that the inclusion of the blackbody to the flux spectra has minor effect on the time-lag fit, as shown in Figure \ref{fig:3models}. This could suggest that the blackbody does not contribute significantly to the correlated variability and thus the time lags. On the other hand, the time-lag residuals near 1~keV motivate that the Gaussian added to the flux spectra may also be required to describe the lags.

It has been found previously some physical parameter estimates inferred from the time lags do not agree with those inferred from the flux spectra, especially the height of the corona \citep{2021ApJ...910L...3W}. When we model the flux spectra independently, the fit parameters are consistent with those found when we include the time lag information, likely due to the fit statistics being dominated by the flux spectra. Fitting the time lags on their own comes with complications in such a high-dimensional parameter space as a result of the large uncertainties on the high-frequency NuSTAR time lags, which are exacerbated by the use of Gaussian processes. As a result, we freeze the temperature to the value from fitting the flux spectra. However, we still find statistically successful solutions with minimal height and very low inclination in which the model entirely smears out the Fe K feature when trying to fit the soft excess. Given that the Fe K lag has been detected in previous observations of this source \citep{2013MNRAS.434.1129K}, we ignore these solutions. Doing so results in a best-fit source height that is lower by roughly a factor of 3 than the value found from simultaneously fitting the lags with the flux spectra, which is a minor discrepancy given the uncertainties ($3.6^{+4.0}_{-1.6} R_g$). This discrepancy is smaller than that found previously by \cite{2021ApJ...910L...3W}, who found this discrepancy in source height to be as large as a factor of 10. An independent fit to the time lags gives an inclination, photon index, and disk electron density consistent with that when fitting the data simultaneously with the additional components. The model requires a maximal iron abundance in addition to lower boost and ionization to counteract the smearing of the reverberation features.

\section{Conclusions} \label{sec:conclusions}
We have simultaneously modeled the time lags and flux spectra of Ark~564 using relativistic reverberation model \textsc{reltrans}. Here are our main results: 
\begin{enumerate}
    \item We present the first results on a long 410~ks NuSTAR observation of Ark~564, simultaneous with two new 115~ks XMM-Newton observations.
    
    \item We implement multi-task learning for the first time for AGN timing, using all of our lightcurves to train a single Gaussian process model for the observed variability in each energy band.
    
    \item We detect the high-frequency time lags and attribute them to reverberation, and model them with the general relativistic ray-tracing model \textsc{reltrans}. 

    \item Through simultaneous modeling of both the XMM-Newton and NuSTAR flux spectra with the lag-energy spectra, we constrain the black hole mass to be $2.3^{+2.6}_{-1.3} \times 10^6M_\odot$. 
    
    \item An independent fit to the time lags gives an inclination, photon index, and disk electron density consistent with that when fitting the lags simultaneously with flux spectra.
    
    \item We find that modeling the soft excess requires additional components, and we speculate that the complexity in modeling the soft excess is related to the high-Eddington nature of this source. This motivates future, intensive analysis, including modeling of the warm absorbers/outflows. 
\end{enumerate}

\begin{acknowledgments}
E.K., G.M., and J.A.G. acknowledge support from NASA grant 80NSSC20K0575. J.J. acknowledges support from the Leverhulme Trust, the Isaac Newton Trust and St Edmund's College, University of Cambridge. A.I. acknowledges support from the Royal Society. G.M. and J.A.G. acknowledge support from NASA grant 80NSSC19K1020. J.A.G. also acknowledges support from the Alexander von Humboldt Foundation. This work was partially supported under NASA contract No. NNG08FD60C and made use of data from the NuSTAR mission, a project led by the California Institute of Technology, managed by the Jet Propulsion Laboratory, and funded by the National Aeronautics and Space Administration. We thank the NuSTAR Operations, Software, and Calibration teams for support with the execution and analysis of these observations. This research has made use of the NuSTAR Data Analysis Software (\textsc{nustardas}), jointly developed by the ASI Science Data Center (ASDC, Italy) and the California Institute of Technology (USA). M.K. acknowledges support through a NWO (Nederlandse Organisatie voor Wetenschappelijk Onderzoek) Spinoza grant. C.S.R. thanks the UK Science and Technology Facilities Council (STFC) for support under the Consolidated Grant ST/S000623/1, as well as the European Research Council (ERC) for support under the European Union’s Horizon 2020 research and innovation programme (grant 834203).
\end{acknowledgments}

\appendix
\setcounter{figure}{0}
\renewcommand{\thefigure}{A\arabic{figure}}
\begin{figure}[h]
    \centering
    \includegraphics[width=0.6\linewidth]{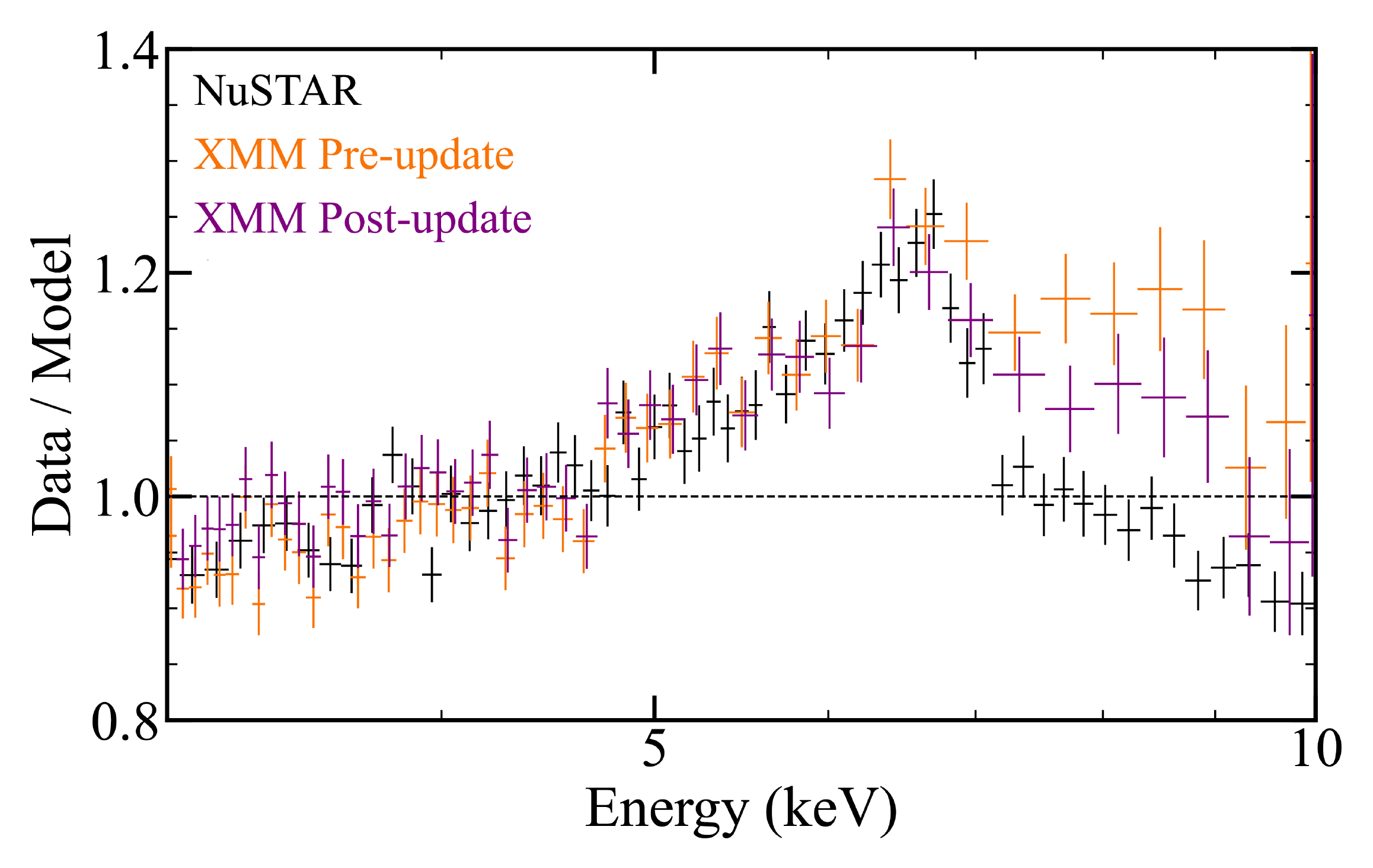}
    \caption{The new NuSTAR spectra (black) and XMM-Newton spectra before and after the release of the current calibration file, including the correction to the effective area (orange and purple, respectively) fit to a power-law. Without the correction, the XMM-Newton spectra exceeds the NuSTAR spectra above roughly 7~keV by $\sim$20$\%$. This discrepancy nearly halves with the effective area correction.}
    \label{fig:7_10_issue}
\end{figure}

\begin{figure}[h] 
    \centering
    \includegraphics[width=0.53\linewidth]{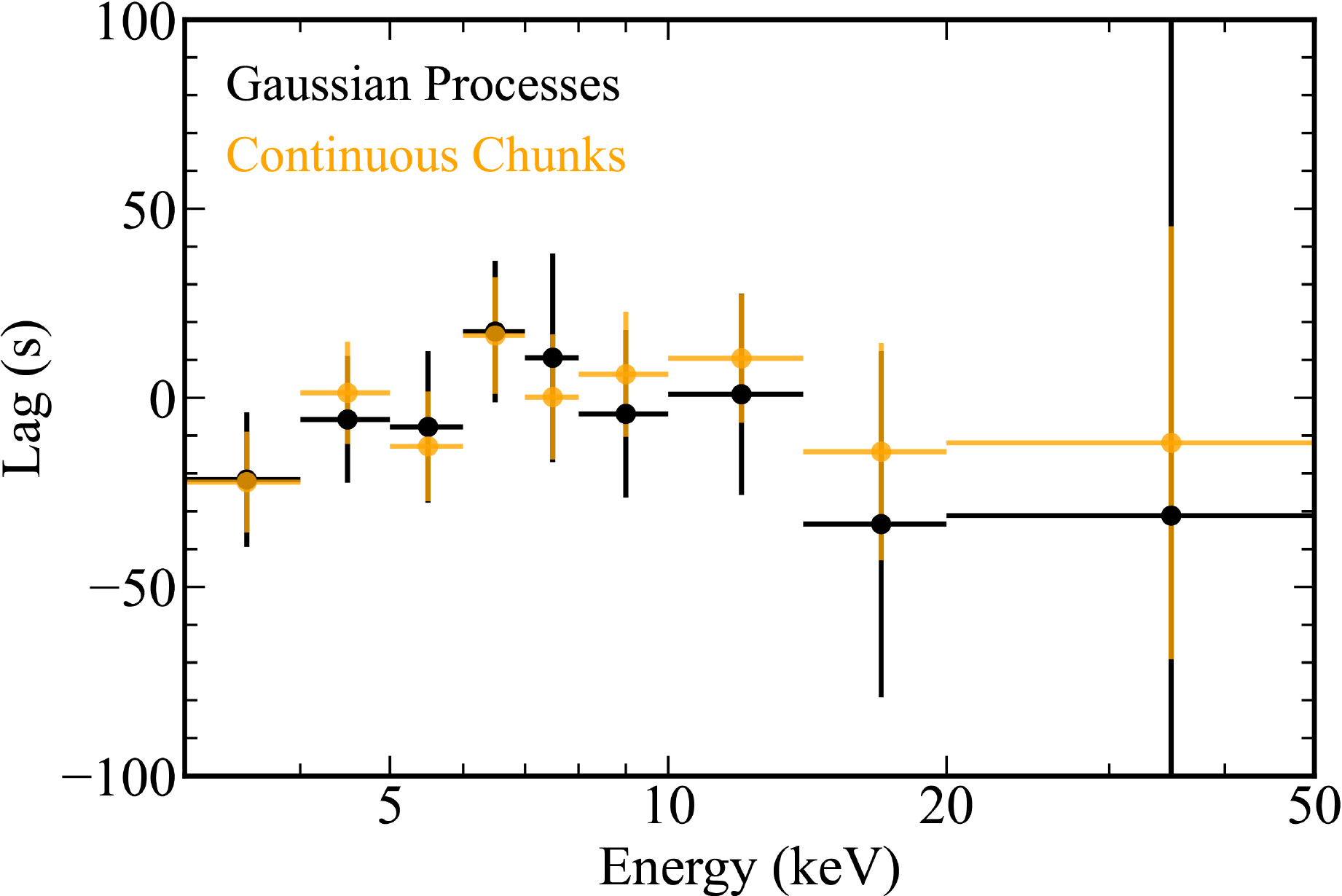}
    \caption{Lag-energy spectra ($0.3-0.9$~mHz) computed from the NuSTAR data using Gaussian processes to generate realizations that include data in the low-Earth orbit gaps (black) versus using the immediately available, continuous chunks (orange). Both approaches give comparable results, but Gaussian processes result in slightly larger uncertainties.}
    \label{fig:gp_vs_chunk}
\end{figure}

\begin{deluxetable}{lcc}[ht!]
\tablecaption{Median and 1$\sigma$-uncertainties on the fit parameters for fitting the flux spectra and lag-energy spectra simultaneously from 2--40~keV with \textsc{reltrans} alone, versus our best 0.5--40~keV fit, including our additional blackbody and Gaussian components \label{tab:comp_table}.}
\tablewidth{0pt}
\tablehead{\colhead{Parameter} & 0.5--40~keV & 2--40~keV}
\startdata
$h$ $[R_g]$  & $9.6^{+0.7}_{-0.7}$ & $9.7^{+1.0}_{-1.0}$ \\
$i$ $[\text{deg}]$ & $37.0^{+0.9}_{-1.0}$ & $38.2^{+1.2}_{-1.2}$ \\
$\Gamma $ & $2.48^{+0.01}_{-0.01}$ &  $2.47^{+0.01}_{-0.02}$\\
$\log(\xi$ $[\text{erg cm/s}])$& $3.25^{+0.03}_{-0.04}$&$3.28^{+0.09}_{-0.08}$\\
$A_{\text{Fe}}/A_{\text{Fe}_\odot}$ & $2.3^{+0.5}_{-0.4}$ & $1.7^{+0.6}_{-0.4}$\\
$\log(n_e$ [cm$^{-3}$]) & $15.1^{+0.2}_{-0.1}$& $15.3^{+0.4}_{-0.2}$\\
$1/\mathcal{B}$ (Boost) & $0.50^{+0.03}_{-0.03}$&$0.51^{+0.07}_{-0.05}$\\
$kT_e$ $[\text{keV}]$  & $14.7^{+1.0}_{-0.9}$& $15.3^{+1.1}_{-1.0}$\\
$\text{Mass}$ $ [10^6 \text{M}_\odot] $ & $2.3^{+2.6}_{-1.3}$ & $3.5^{+3.3}_{-2.3}$\\
$\chi^2_\nu/\nu$  & 1.08/3062& 1.01/2763
\enddata
\end{deluxetable}

\newpage
\bibliography{reference}{}

\begin{thebibliography}{}
\expandafter\ifx\csname natexlab\endcsname\relax\def\natexlab#1{#1}\fi
\providecommand{\url}[1]{\href{#1}{#1}}
\providecommand{\dodoi}[1]{doi:~\href{http://doi.org/#1}{\nolinkurl{#1}}}
\providecommand{\doeprint}[1]{\href{http://ascl.net/#1}{\nolinkurl{http://ascl.net/#1}}}
\providecommand{\doarXiv}[1]{\href{https://arxiv.org/abs/#1}{\nolinkurl{https://arxiv.org/abs/#1}}}

\bibitem[{{Alston} {et~al.}(2020){Alston}, {Fabian}, {Kara}, {Parker},
  {Dovciak}, {Pinto}, {Jiang}, {Middleton}, {Miniutti}, {Walton}, {Wilkins},
  {Buisson}, {Caballero-Garcia}, {Cackett}, {De Marco}, {Gallo}, {Lohfink},
  {Reynolds}, {Uttley}, {Young}, \& {Zogbhi}}]{2020NatAs...4..597A}
{Alston}, W.~N., {Fabian}, A.~C., {Kara}, E., {et~al.} 2020, Nature Astronomy,
  4, 597, \dodoi{10.1038/s41550-019-1002-x}

\bibitem[{{Ar{\'e}valo} \& {Uttley}(2006)}]{2006MNRAS.367..801A}
{Ar{\'e}valo}, P., \& {Uttley}, P. 2006, \mnras, 367, 801,
  \dodoi{10.1111/j.1365-2966.2006.09989.x}

\bibitem[{{Begelman} \& {Meier}(1982)}]{1982ApJ...253..873B}
{Begelman}, M.~C., \& {Meier}, D.~L. 1982, \apj, 253, 873,
  \dodoi{10.1086/159688}

\bibitem[{{Bian} \& {Zhao}(2003)}]{2003ApJ...591..733B}
{Bian}, W., \& {Zhao}, Y. 2003, \apj, 591, 733, \dodoi{10.1086/375414}

\bibitem[{{Botte} {et~al.}(2004){Botte}, {Ciroi}, {Rafanelli}, \& {Di
  Mille}}]{2004AJ....127.3168B}
{Botte}, V., {Ciroi}, S., {Rafanelli}, P., \& {Di Mille}, F. 2004, \aj, 127,
  3168, \dodoi{10.1086/420803}

\bibitem[{Brandt {et~al.}(1997)Brandt, Mathur, \&
  Elvis}]{10.1093/mnras/285.3.L25}
Brandt, W.~N., Mathur, S., \& Elvis, M. 1997, Monthly Notices of the Royal
  Astronomical Society, 285, L25, \dodoi{10.1093/mnras/285.3.L25}

\bibitem[{{Brewer} \& {Stello}(2009)}]{2009MNRAS.395.2226B}
{Brewer}, B.~J., \& {Stello}, D. 2009, \mnras, 395, 2226,
  \dodoi{10.1111/j.1365-2966.2009.14679.x}

\bibitem[{{Cackett} {et~al.}(2021){Cackett}, {Bentz}, \&
  {Kara}}]{2021iSci...24j2557C}
{Cackett}, E.~M., {Bentz}, M.~C., \& {Kara}, E. 2021, iScience, 24, 102557,
  \dodoi{10.1016/j.isci.2021.102557}

\bibitem[{{Cackett} {et~al.}(2014){Cackett}, {Zoghbi}, {Reynolds}, {Fabian},
  {Kara}, {Uttley}, \& {Wilkins}}]{2014MNRAS.438.2980C}
{Cackett}, E.~M., {Zoghbi}, A., {Reynolds}, C., {et~al.} 2014, \mnras, 438,
  2980, \dodoi{10.1093/mnras/stt2424}

\bibitem[{Caruana(1997)}]{10.1023/A:1007379606734}
Caruana, R. 1997, Mach. Learn., 28, 41–75, \dodoi{10.1023/A:1007379606734}

\bibitem[{{Chainakun} {et~al.}(2016){Chainakun}, {Young}, \&
  {Kara}}]{2016MNRAS.460.3076C}
{Chainakun}, P., {Young}, A.~J., \& {Kara}, E. 2016, \mnras, 460, 3076,
  \dodoi{10.1093/mnras/stw1105}

\bibitem[{{Czekala} {et~al.}(2017){Czekala}, {Mandel}, {Andrews}, {Dittmann},
  {Ghosh}, {Montet}, \& {Newton}}]{2017ApJ...840...49C}
{Czekala}, I., {Mandel}, K.~S., {Andrews}, S.~M., {et~al.} 2017, \apj, 840, 49,
  \dodoi{10.3847/1538-4357/aa6aab}

\bibitem[{Dauser {et~al.}(2012)Dauser, Svoboda, Schartel, Wilms, Dovčiak,
  Ehle, Karas, Santos-Lleó, \& Marshall}]{10.1111/j.1365-2966.2011.20356.x}
Dauser, T., Svoboda, J., Schartel, N., {et~al.} 2012, Monthly Notices of the
  Royal Astronomical Society, 422, 1914,
  \dodoi{10.1111/j.1365-2966.2011.20356.x}

\bibitem[{{De Marco} {et~al.}(2013){De Marco}, {Ponti}, {Cappi}, {Dadina},
  {Uttley}, {Cackett}, {Fabian}, \& {Miniutti}}]{2013MNRAS.431.2441D}
{De Marco}, B., {Ponti}, G., {Cappi}, M., {et~al.} 2013, \mnras, 431, 2441,
  \dodoi{10.1093/mnras/stt339}

\bibitem[{{Denney} {et~al.}(2009){Denney}, {Watson}, {Peterson}, {Pogge},
  {Atlee}, {Bentz}, {Bird}, {Brokofsky}, {Comins}, {Dietrich}, {Doroshenko},
  {Eastman}, {Efimov}, {Gaskell}, {Hedrick}, {Klimanov}, {Klimek}, {Kruse},
  {Lamb}, {Leighly}, {Minezaki}, {Nazarov}, {Petersen}, {Peterson},
  {Poindexter}, {Schlesinger}, {Sakata}, {Sergeev}, {Tobin}, {Unterborn},
  {Vestergaard}, {Watkins}, \& {Yoshii}}]{2009ApJ...702.1353D}
{Denney}, K.~D., {Watson}, L.~C., {Peterson}, B.~M., {et~al.} 2009, \apj, 702,
  1353, \dodoi{10.1088/0004-637X/702/2/1353}

\bibitem[{{Earnshaw} \& {Roberts}(2017)}]{2017MNRAS.467.2690E}
{Earnshaw}, H.~M., \& {Roberts}, T.~P. 2017, \mnras, 467, 2690,
  \dodoi{10.1093/mnras/stx308}

\bibitem[{{Fabian} {et~al.}(2000){Fabian}, {Iwasawa}, {Reynolds}, \&
  {Young}}]{2000PASP..112.1145F}
{Fabian}, A.~C., {Iwasawa}, K., {Reynolds}, C.~S., \& {Young}, A.~J. 2000,
  \pasp, 112, 1145, \dodoi{10.1086/316610}

\bibitem[{{Fabian} {et~al.}(2009){Fabian}, {Zoghbi}, {Ross}, {Uttley}, {Gallo},
  {Brandt}, {Blustin}, {Boller}, {Caballero-Garcia}, {Larsson}, {Miller},
  {Miniutti}, {Ponti}, {Reis}, {Reynolds}, {Tanaka}, \&
  {Young}}]{2009Natur.459..540F}
{Fabian}, A.~C., {Zoghbi}, A., {Ross}, R.~R., {et~al.} 2009, \nat, 459, 540,
  \dodoi{10.1038/nature08007}

\bibitem[{Foreman-Mackey(2016)}]{corner}
Foreman-Mackey, D. 2016, The Journal of Open Source Software, 1, 24,
  \dodoi{10.21105/joss.00024}

\bibitem[{{Garc{\'\i}a} {et~al.}(2013){Garc{\'\i}a}, {Dauser}, {Reynolds},
  {Kallman}, {McClintock}, {Wilms}, \& {Eikmann}}]{2013ApJ...768..146G}
{Garc{\'\i}a}, J., {Dauser}, T., {Reynolds}, C.~S., {et~al.} 2013, \apj, 768,
  146, \dodoi{10.1088/0004-637X/768/2/146}

\bibitem[{Garcia {et~al.}(2015)Garcia, Steiner, Mcclintock, Remillard,
  Grinberg, \& Dauser}]{garcia_2015}
Garcia, J., Steiner, J., Mcclintock, J., {et~al.} 2015, The Astrophysical
  Journal, 813, \dodoi{10.1088/0004-637X/813/2/84}

\bibitem[{{Garcia} {et~al.}(2018){Garcia}, {Fabian}, {Grinberg}, {Tomsick},
  {Dauser}, {Wilms}, {Harrison}, {Steiner}, {Sridhar}, {Connors}, \&
  {Wang-Ji}}]{2018cosp...42E1167G}
{Garcia}, J., {Fabian}, A., {Grinberg}, V., {et~al.} 2018, in 42nd COSPAR
  Scientific Assembly, Vol.~42, E1.4--11--18

\bibitem[{{Garc{\'\i}a} {et~al.}(2019){Garc{\'\i}a}, {Kara}, {Walton},
  {Beuchert}, {Dauser}, {Gatuzz}, {Balokovic}, {Steiner}, {Tombesi}, {Connors},
  {Kallman}, {Harrison}, {Fabian}, {Wilms}, {Stern}, {Lanz}, {Ricci}, \&
  {Ballantyne}}]{2019ApJ...871...88G}
{Garc{\'\i}a}, J.~A., {Kara}, E., {Walton}, D., {et~al.} 2019, \apj, 871, 88,
  \dodoi{10.3847/1538-4357/aaf739}

\bibitem[{{George} \& {Fabian}(1991)}]{1991MNRAS.249..352G}
{George}, I.~M., \& {Fabian}, A.~C. 1991, \mnras, 249, 352,
  \dodoi{10.1093/mnras/249.2.352}

\bibitem[{{Giustini} {et~al.}(2015){Giustini}, {Turner}, {Reeves}, {Miller},
  {Legg}, {Kraemer}, \& {George}}]{2015A&A...577A...8G}
{Giustini}, M., {Turner}, T.~J., {Reeves}, J.~N., {et~al.} 2015, \aap, 577, A8,
  \dodoi{10.1051/0004-6361/201425280}

\bibitem[{Griffiths {et~al.}(2021)Griffiths, Jiang, Buisson, Wilkins, Gallo,
  Ingram, Lee, Grupe, Kara, Parker, \& et~al.}]{Griffiths_2021}
Griffiths, R.-R., Jiang, J., Buisson, D. J.~K., {et~al.} 2021, The
  Astrophysical Journal, 914, 144, \dodoi{10.3847/1538-4357/abfa9f}

\bibitem[{Gupta {et~al.}(2013)Gupta, Mathur, Krongold, \&
  Nicastro}]{Gupta:2013in}
Gupta, A., Mathur, S., Krongold, Y., \& Nicastro, F. 2013, Astrophys. J., 772,
  66, \dodoi{10.1088/0004-637X/772/1/66}

\bibitem[{{Ingram} {et~al.}(2019){Ingram}, {Mastroserio}, {Dauser},
  {Hovenkamp}, {van der Klis}, \& {Garc{\'\i}a}}]{2019MNRAS.488..324I}
{Ingram}, A., {Mastroserio}, G., {Dauser}, T., {et~al.} 2019, \mnras, 488, 324,
  \dodoi{10.1093/mnras/stz1720}

\bibitem[{Ingram {et~al.}(2019)Ingram, Mastroserio, Dauser, Hovenkamp,
  van der Klis, \& García}]{reltrans}
Ingram, A., Mastroserio, G., Dauser, T., {et~al.} 2019, Monthly Notices of the
  Royal Astronomical Society, 488, 324, \dodoi{10.1093/mnras/stz1720}

\bibitem[{{Jiang} {et~al.}(2019){Jiang}, {Fabian}, {Dauser}, {Gallo},
  {Garc{\'\i}a}, {Kara}, {Parker}, {Tomsick}, {Walton}, \&
  {Reynolds}}]{density}
{Jiang}, J., {Fabian}, A.~C., {Dauser}, T., {et~al.} 2019, \mnras, 489, 3436,
  \dodoi{10.1093/mnras/stz2326}

\bibitem[{{Jiang} {et~al.}(2014){Jiang}, {Stone}, \&
  {Davis}}]{2014ApJ...796..106J}
{Jiang}, Y.-F., {Stone}, J.~M., \& {Davis}, S.~W. 2014, \apj, 796, 106,
  \dodoi{10.1088/0004-637X/796/2/106}

\bibitem[{{Kalberla} {et~al.}(2005){Kalberla}, {Burton}, {Hartmann}, {Arnal},
  {Bajaja}, {Morras}, \& {P{\"o}ppel}}]{2005A&A...440..775K}
{Kalberla}, P.~M.~W., {Burton}, W.~B., {Hartmann}, D., {et~al.} 2005, \aap,
  440, 775, \dodoi{10.1051/0004-6361:20041864}

\bibitem[{{Kara} {et~al.}(2016){Kara}, {Alston}, {Fabian}, {Cackett}, {Uttley},
  {Reynolds}, \& {Zoghbi}}]{2016MNRAS.462..511K}
{Kara}, E., {Alston}, W.~N., {Fabian}, A.~C., {et~al.} 2016, \mnras, 462, 511,
  \dodoi{10.1093/mnras/stw1695}

\bibitem[{{Kara} {et~al.}(2013){Kara}, {Fabian}, {Cackett}, {Uttley},
  {Wilkins}, \& {Zoghbi}}]{2013MNRAS.434.1129K}
{Kara}, E., {Fabian}, A.~C., {Cackett}, E.~M., {et~al.} 2013, \mnras, 434,
  1129, \dodoi{10.1093/mnras/stt1055}

\bibitem[{Kara {et~al.}(2017)Kara, García, Lohfink, Fabian, Reynolds, Tombesi,
  \& Wilkins}]{Kara_2017}
Kara, E., García, J.~A., Lohfink, A., {et~al.} 2017, Monthly Notices of the
  Royal Astronomical Society, 468, 3489–3498, \dodoi{10.1093/mnras/stx792}

\bibitem[{{Kelly} {et~al.}(2014){Kelly}, {Becker}, {Sobolewska},
  {Siemiginowska}, \& {Uttley}}]{2014ApJ...788...33K}
{Kelly}, B.~C., {Becker}, A.~C., {Sobolewska}, M., {Siemiginowska}, A., \&
  {Uttley}, P. 2014, \apj, 788, 33, \dodoi{10.1088/0004-637X/788/1/33}

\bibitem[{{Kotov} {et~al.}(2001){Kotov}, {Churazov}, \&
  {Gilfanov}}]{2001MNRAS.327..799K}
{Kotov}, O., {Churazov}, E., \& {Gilfanov}, M. 2001, \mnras, 327, 799,
  \dodoi{10.1046/j.1365-8711.2001.04769.x}

\bibitem[{{Lan{\v{c}}ov{\'a}} {et~al.}(2019){Lan{\v{c}}ov{\'a}}, {Abarca},
  {Klu{\'z}niak}, {Wielgus}, {Saḑowski}, {Narayan}, {Schee}, {T{\"o}r{\"o}k},
  \& {Abramowicz}}]{2019ApJ...884L..37L}
{Lan{\v{c}}ov{\'a}}, D., {Abarca}, D., {Klu{\'z}niak}, W., {et~al.} 2019,
  \apjl, 884, L37, \dodoi{10.3847/2041-8213/ab48f5}

\bibitem[{{Legg} {et~al.}(2012){Legg}, {Miller}, {Turner}, {Giustini},
  {Reeves}, \& {Kraemer}}]{2012ApJ...760...73L}
{Legg}, E., {Miller}, L., {Turner}, T.~J., {et~al.} 2012, \apj, 760, 73,
  \dodoi{10.1088/0004-637X/760/1/73}

\bibitem[{{Marconi} {et~al.}(2004){Marconi}, {Risaliti}, {Gilli}, {Hunt},
  {Maiolino}, \& {Salvati}}]{2004MNRAS.351..169M}
{Marconi}, A., {Risaliti}, G., {Gilli}, R., {et~al.} 2004, \mnras, 351, 169,
  \dodoi{10.1111/j.1365-2966.2004.07765.x}

\bibitem[{{Masterson} {et~al.}(2022){Masterson}, {Kara}, {Ricci},
  {Garc{\'\i}a}, {Fabian}, {Pinto}, {Kosec}, {Remillard}, {Loewenstein},
  {Trakhtenbrot}, \& {Arcavi}}]{2022arXiv220605140M}
{Masterson}, M., {Kara}, E., {Ricci}, C., {et~al.} 2022, arXiv e-prints,
  arXiv:2206.05140.
\newblock \doarXiv{2206.05140}

\bibitem[{{Mastroserio} {et~al.}(2018){Mastroserio}, {Ingram}, \& {van der
  Klis}}]{2018MNRAS.475.4027M}
{Mastroserio}, G., {Ingram}, A., \& {van der Klis}, M. 2018, \mnras, 475, 4027,
  \dodoi{10.1093/mnras/sty075}

\bibitem[{{Mastroserio} {et~al.}(2020){Mastroserio}, {Ingram}, \& {van der
  Klis}}]{2020MNRAS.498.4971M}
---. 2020, \mnras, 498, 4971, \dodoi{10.1093/mnras/staa2735}

\bibitem[{{Mastroserio} {et~al.}(2021){Mastroserio}, {Ingram}, {Wang},
  {Garc{\'\i}a}, {van der Klis}, {Cavecchi}, {Connors}, {Dauser}, {Harrison},
  {Kara}, {K{\"o}nig}, \& {Lucchini}}]{2021MNRAS.507...55M}
{Mastroserio}, G., {Ingram}, A., {Wang}, J., {et~al.} 2021, \mnras, 507, 55,
  \dodoi{10.1093/mnras/stab2056}

\bibitem[{{Matsumoto} {et~al.}(2004){Matsumoto}, {Leighly}, \&
  {Marshall}}]{2004ApJ...603..456M}
{Matsumoto}, C., {Leighly}, K.~M., \& {Marshall}, H.~L. 2004, \apj, 603, 456,
  \dodoi{10.1086/381666}

\bibitem[{{McAllister} {et~al.}(2017){McAllister}, {Littlefair}, {Dhillon},
  {Marsh}, {Ashley}, {Bours}, {Breedt}, {Hardy}, {Hermes}, {Kengkriangkrai},
  {Kerry}, {Rattanasoon}, \& {Sahman}}]{2017MNRAS.464.1353M}
{McAllister}, M.~J., {Littlefair}, S.~P., {Dhillon}, V.~S., {et~al.} 2017,
  \mnras, 464, 1353, \dodoi{10.1093/mnras/stw2417}

\bibitem[{{McHardy} {et~al.}(2007){McHardy}, {Ar{\'e}valo}, {Uttley},
  {Papadakis}, {Summons}, {Brinkmann}, \& {Page}}]{2007MNRAS.382..985M}
{McHardy}, I.~M., {Ar{\'e}valo}, P., {Uttley}, P., {et~al.} 2007, \mnras, 382,
  985, \dodoi{10.1111/j.1365-2966.2007.12411.x}

\bibitem[{{Miller} {et~al.}(2010){Miller}, {Turner}, {Reeves}, {Lobban},
  {Kraemer}, \& {Crenshaw}}]{2010MNRAS.403..196M}
{Miller}, L., {Turner}, T.~J., {Reeves}, J.~N., {et~al.} 2010, \mnras, 403,
  196, \dodoi{10.1111/j.1365-2966.2009.16149.x}

\bibitem[{{Niko{\l}ajuk} {et~al.}(2009){Niko{\l}ajuk}, {Czerny}, \&
  {Gurynowicz}}]{2009MNRAS.394.2141N}
{Niko{\l}ajuk}, M., {Czerny}, B., \& {Gurynowicz}, P. 2009, \mnras, 394, 2141,
  \dodoi{10.1111/j.1365-2966.2009.14478.x}

\bibitem[{{Pinto} {et~al.}(2021){Pinto}, {Soria}, {Walton}, {D'A{\`\i}},
  {Pintore}, {Kosec}, {Alston}, {Fuerst}, {Middleton}, {Roberts}, {Del Santo},
  {Barret}, {Ambrosi}, {Robba}, {Earnshaw}, \& {Fabian}}]{2021MNRAS.505.5058P}
{Pinto}, C., {Soria}, R., {Walton}, D.~J., {et~al.} 2021, \mnras, 505, 5058,
  \dodoi{10.1093/mnras/stab1648}

\bibitem[{Rasmussen \& Williams(2006)}]{10.5555/1162254}
Rasmussen, C.~E., \& Williams, C. K.~I. 2006, Gaussian Processes for Machine
  Learning (Adaptive Computation and Machine Learning) (The MIT Press)

\bibitem[{{Ross} \& {Fabian}(2005)}]{2005MNRAS.358..211R}
{Ross}, R.~R., \& {Fabian}, A.~C. 2005, \mnras, 358, 211,
  \dodoi{10.1111/j.1365-2966.2005.08797.x}

\bibitem[{{Sadowski}(2011)}]{2011arXiv1108.0396S}
{Sadowski}, A. 2011, arXiv e-prints, arXiv:1108.0396.
\newblock \doarXiv{1108.0396}

\bibitem[{{Sadowski} {et~al.}(2014){Sadowski}, {Narayan}, {McKinney}, \&
  {Tchekhovskoy}}]{2014MNRAS.439..503S}
{Sadowski}, A., {Narayan}, R., {McKinney}, J.~C., \& {Tchekhovskoy}, A. 2014,
  \mnras, 439, 503, \dodoi{10.1093/mnras/stt2479}

\bibitem[{{Shemmer} {et~al.}(2001){Shemmer}, {Romano}, {Bertram}, {Brinkmann},
  {Collier}, {Crowley}, {Detsis}, {Filippenko}, {Gaskell}, {George}, {Gliozzi},
  {Hiller}, {Jewell}, {Kaspi}, {Klimek}, {Lannon}, {Li}, {Martini}, {Mathur},
  {Negoro}, {Netzer}, {Papadakis}, {Papamastorakis}, {Peterson}, {Peterson},
  {Pogge}, {Pronik}, {Rumstay}, {Sergeev}, {Sergeeva}, {Stirpe}, {Taylor},
  {Treffers}, {Turner}, {Uttley}, {Vestergaard}, {von Braun}, {Wagner}, \&
  {Zheng}}]{2001ApJ...561..162S}
{Shemmer}, O., {Romano}, P., {Bertram}, R., {et~al.} 2001, \apj, 561, 162,
  \dodoi{10.1086/323236}

\bibitem[{{Str\"uder, L.} {et~al.}(2001){Str\"uder, L.}, {Briel, U.}, {Dennerl,
  K.}, {Hartmann, R.}, {Kendziorra, E.}, {Meidinger, N.}, {Pfeffermann, E.},
  {Reppin, C.}, {Aschenbach, B.}, {Bornemann, W.}, {Br\"auninger, H.},
  {Burkert, W.}, {Elender, M.}, {Freyberg, M.}, {Haberl, F.}, {Hartner, G.},
  {Heuschmann, F.}, {Hippmann, H.}, {Kastelic, E.}, {Kemmer, S.}, {Kettenring,
  G.}, {Kink, W.}, {Krause, N.}, {M\"uller, S.}, {Oppitz, A.}, {Pietsch, W.},
  {Popp, M.}, {Predehl, P.}, {Read, A.}, {Stephan, K. H.}, {St\"otter, D.},
  {Tr\"umper, J.}, {Holl, P.}, {Kemmer, J.}, {Soltau, H.}, {St\"otter, R.},
  {Weber, U.}, {Weichert, U.}, {von Zanthier, C.}, {Carathanassis, D.}, {Lutz,
  G.}, {Richter, R. H.}, {Solc, P.}, {B\"ottcher, H.}, {Kuster, M.}, {Staubert,
  R.}, {Abbey, A.}, {Holland, A.}, {Turner, M.}, {Balasini, M.}, {Bignami, G.
  F.}, {La Palombara, N.}, {Villa, G.}, {Buttler, W.}, {Gianini, F.}, {Lain\'e,
  R.}, {Lumb, D.}, \& {Dhez, P.}}]{epicpn}
{Str\"uder, L.}, {Briel, U.}, {Dennerl, K.}, {et~al.} 2001, A\&A, 365, L18,
  \dodoi{10.1051/0004-6361:20000066}

\bibitem[{{Sunyaev} \& {Truemper}(1979)}]{1979Natur.279..506S}
{Sunyaev}, R.~A., \& {Truemper}, J. 1979, \nat, 279, 506,
  \dodoi{10.1038/279506a0}

\bibitem[{{Taylor} \& {Reynolds}(2018)}]{2018ApJ...855..120T}
{Taylor}, C., \& {Reynolds}, C.~S. 2018, \apj, 855, 120,
  \dodoi{10.3847/1538-4357/aaad63}

\bibitem[{{Timmer} \& {Koenig}(1995)}]{timmer}
{Timmer}, J., \& {Koenig}, M. 1995, \aap, 300, 707

\bibitem[{{Tomsick} {et~al.}(2018){Tomsick}, {Parker}, {Garc{\'\i}a},
  {Yamaoka}, {Barret}, {Chiu}, {Clavel}, {Fabian}, {F{\"u}rst}, {Gandhi},
  {Grinberg}, {Miller}, {Pottschmidt}, \& {Walton}}]{tomsick}
{Tomsick}, J.~A., {Parker}, M.~L., {Garc{\'\i}a}, J.~A., {et~al.} 2018, \apj,
  855, 3, \dodoi{10.3847/1538-4357/aaaab1}

\bibitem[{{Tripathi} {et~al.}(2018){Tripathi}, {Nampalliwar}, {Abdikamalov},
  {Ayzenberg}, {Jiang}, \& {Bambi}}]{2018PhRvD..98b3018T}
{Tripathi}, A., {Nampalliwar}, S., {Abdikamalov}, A.~B., {et~al.} 2018, \prd,
  98, 023018, \dodoi{10.1103/PhysRevD.98.023018}

\bibitem[{{Urquhart} \& {Soria}(2016)}]{2016MNRAS.456.1859U}
{Urquhart}, R., \& {Soria}, R. 2016, \mnras, 456, 1859,
  \dodoi{10.1093/mnras/stv2293}

\bibitem[{{Uttley} {et~al.}(2014){Uttley}, {Cackett}, {Fabian}, {Kara}, \&
  {Wilkins}}]{2014A&ARv..22...72U}
{Uttley}, P., {Cackett}, E.~M., {Fabian}, A.~C., {Kara}, E., \& {Wilkins},
  D.~R. 2014, \aapr, 22, 72, \dodoi{10.1007/s00159-014-0072-0}

\bibitem[{Uttley {et~al.}(2014)Uttley, Cackett, Fabian, Kara, \&
  Wilkins}]{Uttley_2014}
Uttley, P., Cackett, E.~M., Fabian, A.~C., Kara, E., \& Wilkins, D.~R. 2014,
  The Astronomy and Astrophysics Review, 22, \dodoi{10.1007/s00159-014-0072-0}

\bibitem[{Walton {et~al.}(2012)Walton, Nardini, Fabian, Gallo, \&
  Reis}]{10.1093/mnras/sts227}
Walton, D.~J., Nardini, E., Fabian, A.~C., Gallo, L.~C., \& Reis, R.~C. 2012,
  Monthly Notices of the Royal Astronomical Society, 428, 2901,
  \dodoi{10.1093/mnras/sts227}

\bibitem[{{Wang} {et~al.}(2021){Wang}, {Mastroserio}, {Kara}, {Garc{\'\i}a},
  {Ingram}, {Connors}, {van der Klis}, {Dauser}, {Steiner}, {Buisson}, {Homan},
  {Lucchini}, {Fabian}, {Bright}, {Fender}, {Cackett}, \&
  {Remillard}}]{2021ApJ...910L...3W}
{Wang}, J., {Mastroserio}, G., {Kara}, E., {et~al.} 2021, \apjl, 910, L3,
  \dodoi{10.3847/2041-8213/abec79}

\bibitem[{{Wang} \& {Lu}(2001)}]{2001A&A...377...52W}
{Wang}, T., \& {Lu}, Y. 2001, \aap, 377, 52, \dodoi{10.1051/0004-6361:20011071}

\bibitem[{Wilkins(2019)}]{Wilkins_2019}
Wilkins, D.~R. 2019, Monthly Notices of the Royal Astronomical Society, 489,
  1957–1972, \dodoi{10.1093/mnras/stz2269}

\bibitem[{{Willecke Lindberg} {et~al.}(2021){Willecke Lindberg},
  {Huppenkothen}, {Jones}, {Bolin}, {Juric}, {Golkhou}, {Bellm}, {Drake},
  {Graham}, {Laher}, {Mahabal}, {Masci}, {Riddle}, \&
  {Shin}}]{2021arXiv211112596W}
{Willecke Lindberg}, C., {Huppenkothen}, D., {Jones}, R.~L., {et~al.} 2021,
  arXiv e-prints, arXiv:2111.12596.
\newblock \doarXiv{2111.12596}

\bibitem[{{Wilms} {et~al.}(2000){Wilms}, {Allen}, \&
  {McCray}}]{2000ApJ...542..914W}
{Wilms}, J., {Allen}, A., \& {McCray}, R. 2000, \apj, 542, 914,
  \dodoi{10.1086/317016}

\bibitem[{{Xu} {et~al.}(2022){Xu}, {Pinto}, {Kara}, {Masterson}, {Garc{\'\i}a},
  {Fabian}, {Parker}, {Barret}, {Alston}, \& {Cusumano}}]{2022arXiv220406075X}
{Xu}, Y., {Pinto}, C., {Kara}, E., {et~al.} 2022, arXiv e-prints,
  arXiv:2204.06075.
\newblock \doarXiv{2204.06075}

\bibitem[{{Zdziarski} {et~al.}(1996){Zdziarski}, {Johnson}, \&
  {Magdziarz}}]{1996MNRAS.283..193Z}
{Zdziarski}, A.~A., {Johnson}, W.~N., \& {Magdziarz}, P. 1996, \mnras, 283,
  193, \dodoi{10.1093/mnras/283.1.193}

\bibitem[{{Zoghbi} {et~al.}(2020){Zoghbi}, {Kalli}, {Miller}, \&
  {Mizumoto}}]{2020ApJ...893...97Z}
{Zoghbi}, A., {Kalli}, S., {Miller}, J.~M., \& {Mizumoto}, M. 2020, \apj, 893,
  97, \dodoi{10.3847/1538-4357/ab7dc8}

\bibitem[{{Zoghbi} {et~al.}(2013){Zoghbi}, {Reynolds}, \&
  {Cackett}}]{2013ApJ...777...24Z}
{Zoghbi}, A., {Reynolds}, C., \& {Cackett}, E.~M. 2013, \apj, 777, 24,
  \dodoi{10.1088/0004-637X/777/1/24}

\bibitem[{{Zoghbi} {et~al.}(2011){Zoghbi}, {Uttley}, \&
  {Fabian}}]{2011MNRAS.412...59Z}
{Zoghbi}, A., {Uttley}, P., \& {Fabian}, A.~C. 2011, \mnras, 412, 59,
  \dodoi{10.1111/j.1365-2966.2010.17883.x}

\bibitem[{{{\.Z}ycki} {et~al.}(1999){{\.Z}ycki}, {Done}, \&
  {Smith}}]{1999MNRAS.309..561Z}
{{\.Z}ycki}, P.~T., {Done}, C., \& {Smith}, D.~A. 1999, \mnras, 309, 561,
  \dodoi{10.1046/j.1365-8711.1999.02885.x}

\end{thebibliography}
\bibliographystyle{aasjournal}

\end{document}